\documentclass[aps,prl,twocolumn,superscriptaddress,english]{revtex4}
\usepackage{amssymb}
\usepackage{graphicx}
\usepackage{amsmath}
\usepackage{amsthm}
\usepackage{amsfonts}
\usepackage[T1]{fontenc}
\usepackage[latin9]{inputenc}
\usepackage{array}
\usepackage{multirow}
\usepackage{color}
\usepackage{esint}
\usepackage{bm}
\usepackage{color}
\usepackage{bbm}
\usepackage{hyperref}
\usepackage{babel}
\usepackage{titlesec}

\def\Arg{\mathop{\operator@font Arg}\nolimits}

\begin{document}

\global\long\def\id{\mathbbm{1}}
\global\long\def\ui{\mathbbm{i}}
\global\long\def\ud{\mathrm{d}}

\title{Ideal Weyl semimetal with 3D spin-orbit coupled ultracold quantum gas}

\author{Yue-Hui Lu}
\affiliation{International Center for Quantum Materials and School of Physics, Peking University, Beijing, China 100871}
\affiliation{Collaborative Innovation Center of Quantum Matter, Beijing 100871, China}
\author{Bao-Zong Wang}
\affiliation{Shanghai Branch, National Laboratory for Physical Sciences
at Microscale and Department of Modern Physics, University
of Science and Technology of China, Shanghai 201315,
China}
\affiliation{International Center for Quantum Materials and School of Physics, Peking University, Beijing, China 100871}
\author{Xiong-Jun Liu}\thanks{Corresponding author: xiongjunliu@pku.edu.cn}
\affiliation{International Center for Quantum Materials and School of Physics, Peking University, Beijing, China 100871}
\affiliation{Collaborative Innovation Center of Quantum Matter, Beijing 100871, China}
\affiliation{CAS Center for Excellence in Topological Quantum Computation, University of Chinese Academy of Sciences, Beijing 100190, China}


\begin{abstract}
There is an immense effort in search for various types of Weyl semimetals, of which the most fundamental phase consists of the minimal number of i.e. two Weyl points, but is hard to engineer in solids. Here we demonstrate how such fundamental Weyl semimetal can be realized in a maneuverable optical Raman lattice, with which the three-dimensional (3D) spin-orbit (SO) coupling is synthesised for ultracold atoms. In addition, a new novel Weyl phase with coexisting Weyl nodal points and nodal ring is also predicted here, and is shown to be protected by nontrivial linking numbers. We further propose feasible techniques to precisely resolve 3D Weyl band topology through 2D equilibrium and dynamical measurements. 
This work leads to the first realization of the most fundamental Weyl semimetal band and the 3D SO coupling for ultracold quantum gases, which are respectively the significant issues in the condensed matter and ultracold atom physics.
\end{abstract}

\maketitle

{\it Introduction.}--Weyl fermion is a massless particle with definite chirality~\cite{Weyl1929}. While not identified as a fundamental particle in nature, the Weyl fermion can emerge as low-energy quasiparticle in Weyl semimetals, which have been discovered in solids~\cite{xu2015discovery,lv2015experimental,Xu2015NatPhys,hasan2017discovery} and also simulated in meta-materials~\cite{Lu2015,Noh2017,Lu2019PRB}, and generated considerable interests in recent years~\cite{armitage2018weyl}.
In the crystal Fermion doubling ensures that Weyl points come in pairs, with two Weyl nodes in each pair having opposite chiralities~\cite{Nielsen1981,Wan2011,Burkov2011}. Thus the minimal number of Weyl points in a semimetal is two, and such minimal case renders the most fundamental phase in Weyl semimetal family as well as an ideal Weyl semimetal (IWSM)~\cite{IWSM1}. The two Weyl nodes in such IWSM are not symmetry-related, and all basic symmetries in IWSM can be broken~\cite{armitage2018weyl}. 


The IWSM phase has peculiar importance. 
As protected solely by Chern numbers, the two separated Weyl nodes of opposite chiralities in the IWSM cannot be trivially gapped out. This leads to an important consequence that any interacting phase born of IWSM has to be topologically nontrivial, being either gapless with new Weyl points, or gapped with exotic bulk topology. For example, the superconductivity with pairing between two different Weyl cones gives a gapless Weyl superconductor~\cite{BCS1,BCS2,FFLO2,FFLO3}. Further, the Fulde-Ferrell-Larkin-Ovchinnikov pairing order within each Weyl cone in the IWSM renders more exotic gapped phases exhibiting emergent space-time supersymmetry~\cite{Yao2015PRL} or hosting non-Abelian Majorana modes protected by second Chern numbers~\cite{Chan2017}. The similar exotic excitonic phases in the IWSM with repulsive interactions have also been predicted~\cite{Ye2016}. In contrast, an interacting phase generated from Weyl semimetal with four or more Weyl points can be either topological or trivial, since each two Weyl cones of the same chirality can be trivially gapped out with e.g. conventional superconducting pairings~\cite{Qi2014,Kim216,Bitan2017,Wang2017}.

The IWSM is hard to engineer in solids, while the very recent success in observing magnetic Weyl semimetals comes closer to its realization~\cite{Liu2019science,Morali2019science,Belopolskiscience2019}. 
Progresses have been also made in simulating topological states in ultracold atoms~\cite{Jotzu2014,Wu:2016kv,Li:2016tp,Song:2017uf,Song2019,Cooper2019}. However, to achieve the IWSM in ultracold atoms faces challenges not only in realizing 3D spin-orbit (SO) coupling, which is fundamentally different from the currently realized 1D~\cite{Lin_SOC} and 2D SO couplings~\cite{Wu:2016kv,Huang2016,Goldman2014,LZhang2018Review}, but also in measuring 3D phases~\cite{Xu2015,Zhang2015,Wang2016,He:2016,Juan2017,Tran2017,XuPRB2019}. The realization of 3D SO coupling by itself is highly nontrivial but was not successful for ultracold atoms.
In this article, we solve all the challenges and design a minimal scheme to realize 3D SO coupling and IWSM, with a new coexisting Weyl phase being also predicted and shown to be protected by nontrivial linking numbers. We further propose experimental techniques including the virtual slicing technique and dynamical schemes to detect the 3D band topology via 2D measurements, which are of high feasibility in experiment. This work paves the way for the realization and detection of the most fundamental Weyl semimetal phase and the 3D SO coupling for ultracold atoms.

\begin{figure*}
\centering
\includegraphics[width=0.8\textwidth, page=1]{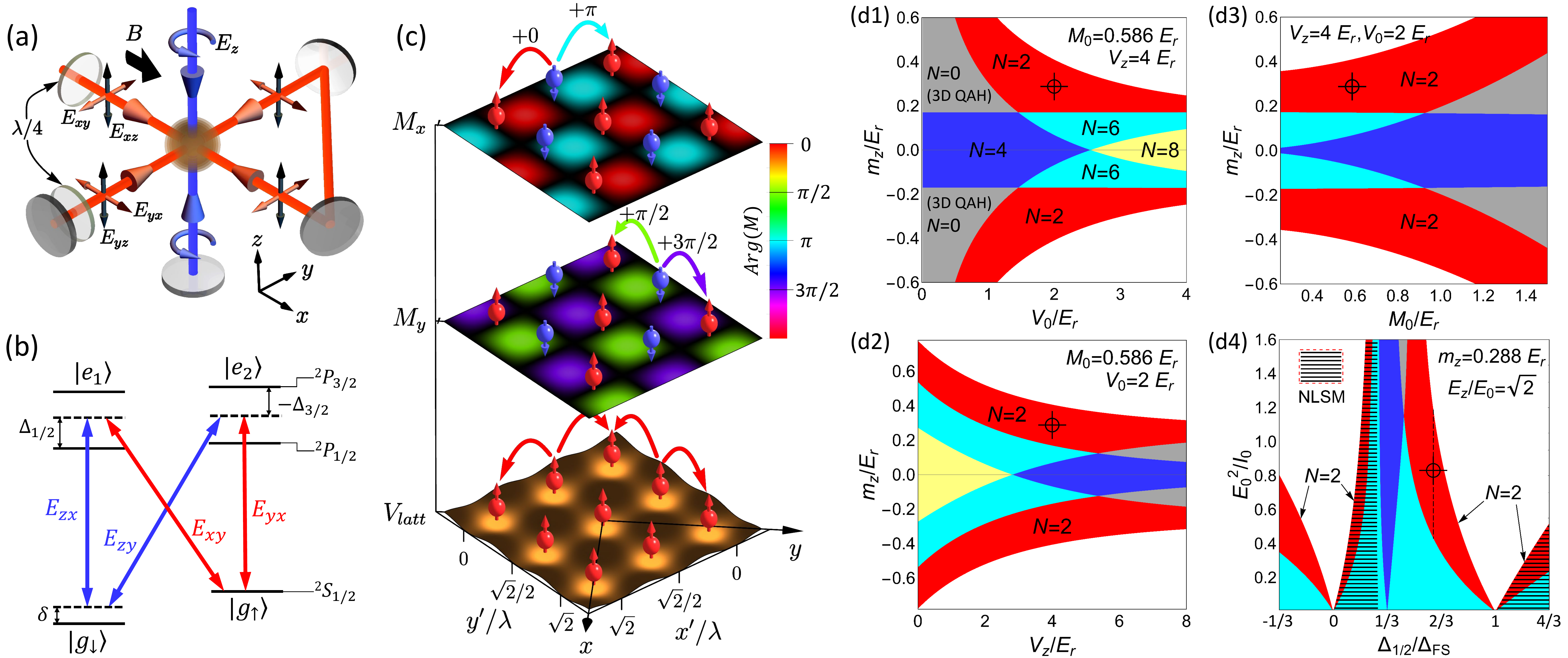}
\caption{\label{fig:setup} (a) {Minimal setup for realization,} with bias magnetic field in $x$ direction. The incident beam ${E}_x$ has $(y,z)$ polarization components (red line), reflected by mirrors. {The orthogonal polarization components acquire $\pi/2$ relative phase once passing through quarter-wave-plate, forming checkerboard lattice in $x'$ and $y'$ axes [bottom layer of (c)].} A standing-wave beam (blue line) forms the lattice in $z$ direction. (b) Configuration for the two-photon Raman transitions, which are typically contributed from the D1 and D2 lines, with detunings $\Delta_{1/2}$ and $-\Delta_{3/2}$, respectively~\cite{SI}. (c) Profiles of the lattice and Raman potentials in $x-y$ plane. (d) Phase diagrams. The parameters $m_z, V_0, V_z$, and $M_0$ represent the effective Zeeman field, lattice depths in $x-y$ plane and $z$ direction, and Raman coupling strength. In (d4) the quantity $I_0=3E_r\Delta_{FS}/(e^2a_0^2)$ for alkali atoms, with $a_0$ and $\Delta_{FS}$ denoting Bohr radius, and the fine-structure splitting between D1 and D2 lines, respectively~\cite{SI}. The topological phases with $N$ ($=2,4,6,8$) Weyl points are colored, {the 3D QAH phase in black areas is fully gapped with each $q_z$-layer being nontrivial, and the white area is trivial with $N=0$}. The zebra-lined areas {denote the novel phase with coexisting Weyl points and nodal ring} (see Supplementary Material~\cite{SI}). The red areas give the IWSM phase with $N=2$. The chosen parameters are marked as the bull's eye on each diagram.}
\end{figure*}

{\it The model.}--We start with the model realization via optical Raman lattice scheme. The
essential ingredients include a {\it configuration-tunable} 3D optical lattice and periodic Raman potentials which have nontrivial relative symmetries with respect to the lattice [Fig.~\ref{fig:setup}(a-c)]. Two Zeeman-split ground hyperfine levels $|g_{\uparrow,\downarrow}\rangle$ are considered.
The laser beam $E_x$ incident from $x$ direction has polarizations in $y\text{{-}}z$ plane, reflected by three mirrors together with $\pi/2$ relative phase induced by two quarter-wave plates between the in-plane ($x$ or $y$) and out-of-plane ($z$) polarization components [Fig.~\ref{fig:setup}(a)]. 
This gives $x$-$y$ plane electric fields: $\bold{E}_{x-y}= 2[E_{xy}(\bold{\hat{y}}\sin k_{0}x - \bold{\hat{x}}\sin k_{0}y)e^{-i\omega_1t} +\bold{\hat{z}}E_{xz}(\cos k_{0}x+\cos k_{0}y)e^{-i\omega_zt}]$, where $E_{\alpha\beta}(\alpha,\beta=x,y,z)$ is the amplitude of the field propagating in the $\alpha$ direction with the $\beta$ polarization, and we have taken $E_{xy}=E_{yx}$ and $E_{xz}=E_{yz}$. The difference of frequencies ($\omega_1$ and $\omega_z$) of the two components $E_{xy}$ and $E_{xz}$ is in the order of MHz, hence the two components have approximately the same wave vector $k_0$. In the $z$ direction, a circularly polarized beam of frequency $\omega_2$ forms standing wave $\bold{E}_z=(E_{zx}\bold{\hat{x}}+iE_{zy}\bold{\hat{y}})\cos k_{0}ze^{-i\omega_2t}$. The two sets of beams generate both lattice and Raman potentials as $\delta\omega=\omega_2-\omega_1$ compensates the Zeeman splitting $\mu_B g_F B/\hbar$ in $|g_{\uparrow,\downarrow}\rangle$ except for a small two-photon detuing $\delta$ [Fig.~\ref{fig:setup}(b)]~\cite{Liu2013,Liu:2014hj,Wu:2016kv,Wang:2018ic,Song:2017uf}. The $E_{xz}$-component of $\bold{E}_{x-y}$ does not contribute to Raman couplings as $\omega_z$ is tuned to have several MHz difference from $\omega_{1,2}$, far detuned from the two-photon Raman resonance.

The lattice configuration is tuned by {the incident components} $E_{xy, xz}$. The in-plane component ($E_{xy}$) solely generates a regular square lattice in $x$-$y$ plane, which deforms to a checkerboard one with neighboring sites along the diagonal $x'=(x-y)/\sqrt{2}$ and $y'=(x+y)/\sqrt{2}$ directions if turning on $E_{xz}$ and tuning to $E_{xy}=E_{xz}=E_0/2$ [Fig.~\ref{fig:setup}(c)]. As we shall see below, this deformed configuration is essential in our realization of the 3D SO coupling and IWSM. The lattice potential
$\mathcal{V}_{latt} \propto (\bold{E}_{x-y}^*\cdot\bold{E}_{x-y}+\bold{E}_z^*\cdot\bold{E}_z)$ takes the form $\mathcal{V}_{latt} = -V_0 \big[\cos^2(k'_{0}x')+\cos^2(k'_{0}y')\big] - V_z \cos^2 k_0z$. Here $k'_{0}=k_0/\sqrt{2}$ and lattice depths $(V_0,V_z)\propto(E_0^2,E_z^2)>0$ for the overall red-detuned regime. Further, the Raman potentials are generated through two-photon processes by the {${E}_{yx}$ (${E}_{xy}$)} and {${E}_{zy}$ (${E}_{zx}$) components} [Fig.~\ref{fig:setup}(b)], giving $\mathcal{M}(\bold{r})=M_x\sigma_x+M_y\sigma_y=M_{0}\cos k_{0}z(\sin k_{0}x\sigma_{x}+\sin k_{0}y\sigma_{y})$, with $M_0 \propto E_z E_0$. We finally reach~\cite{SI}
\begin{eqnarray}\label{Hamiltonian1}
\mathcal{H}=\left[\frac{\bold{p}^2}{2m}+\mathcal{V}_{latt}(\bold{r})\right]\mathbf{1} +\mathcal{M}(\bold{r})+m_z \sigma_z, \ m_z=\frac{\delta}{2}.
\end{eqnarray}
The Raman potentials {$M_{x}(x,z)$ and $M_y(y,z)$ are each in 2D forms in $(x,y,z)$ coordinates. However, they are 3D forms in the checkerboard lattice bases $\mathcal{M}(\bold{r})=\mathcal{M}_{x'} \sigma_x+\mathcal{M}_{y'} \sigma_y$, with $\mathcal{M}_{x'}=\sqrt{2}
M_{0} \sin (k'_{0}x')\cos (k'_{0}y')\cos k_{0}z$ and $\mathcal{M}_{y'}=\mathcal{M}_{x'}(x'\leftrightarrow y')$ being reflection anti-symmetric with respect to each lattice center of $\mathcal{V}_{latt}(\bold{r})$ in the $x'$ and $y'$ directions, respectively, but symmetric in other directions [Fig.~\ref{fig:setup}(c)]. The 3D features of $\mathcal{M}_{x',y'}$ with the nontrivial relative symmetry are essential for realizing the 3D SO coupling and} IWSM. The similar relative symmetry was previously achieved only for 2D quantum anomalous Hall (QAH) model~\cite{Liu:2014hj,Wang:2018ic}.

\begin{figure*}[t]
\centering
\includegraphics[width=0.7\textwidth]{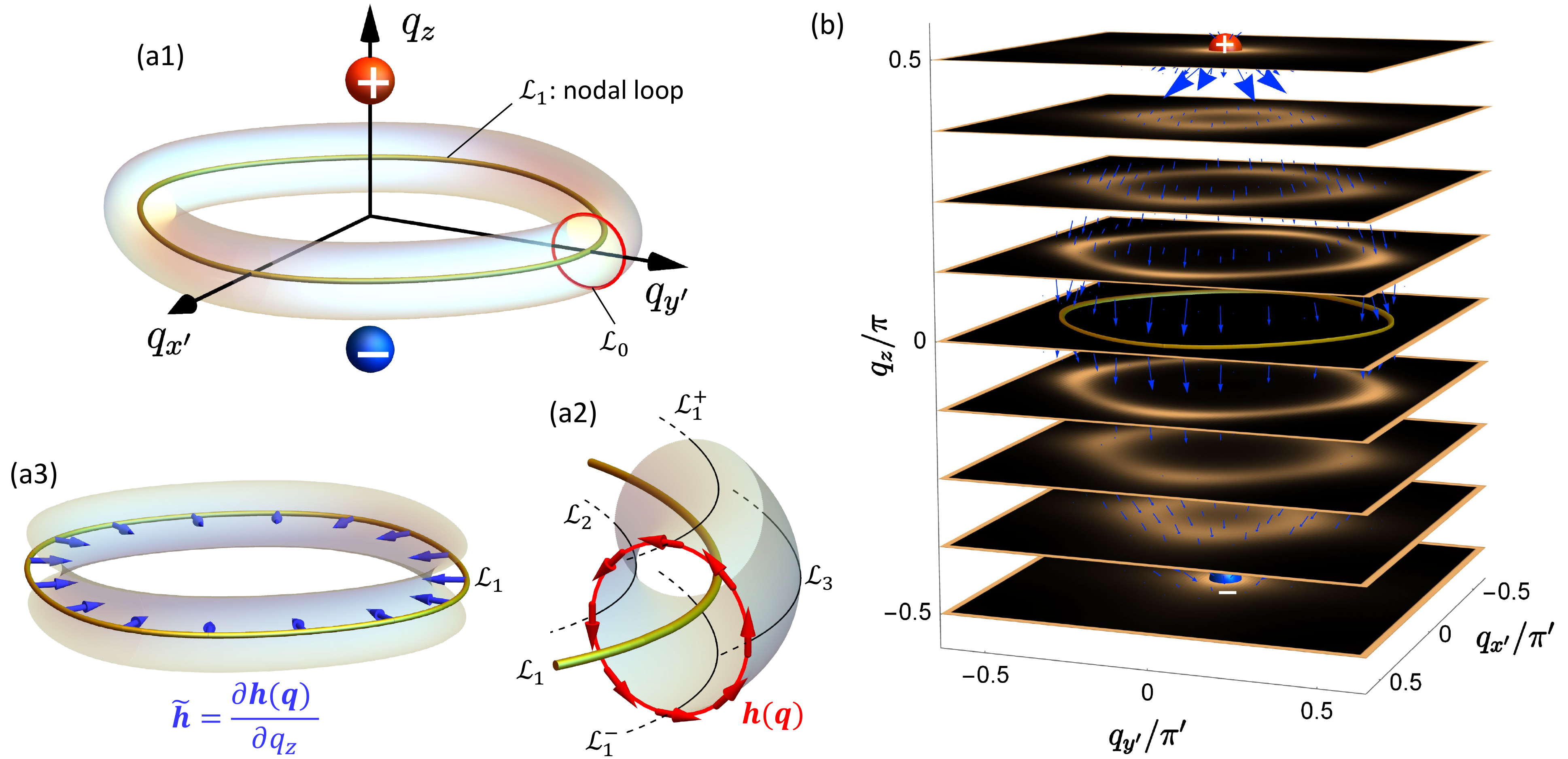}
\caption{\label{SIfig:NLSM} Characterization of the new Weyl phase with coexisting Weyl nodal point and nodal ring. {(a1)} A schematic of the Weyl points with positive and negative charges, respectively, and nodal ring $\mathcal{L}_1$ (also denoted as yellow ring in (a2,a3)). {(a2)} The vector field $\bold{h}(\bold{q})$ on the small loop $\mathcal{L}_0$ that threads the nodal loop, rendering a winding number $w_0 = 1$. {(a3)} The vector field $\tilde{\bold{h}}(q_{x'},q_{y'})$ of the effective Hamiltonian $\tilde{\cal H}(q_{x'},q_{y'}) = \partial{\cal H}(\bold{q})/\partial q_z|_{q_z=0}$ on the nodal loop $\mathcal{L}_1$, rendering a winding number $w_1 = -1$, which together with $w_0$ forms a linking number. {(b)} Berry curvature flux on a series of $q_z$-slices ranging between the two Weyl nodes. 
The parameters are taken as $m_z/t_0 = 4, t_z/t_0 = 1, t_{SO}/t_0 = 0.25$ in the tight-binding model.
}
\end{figure*}

\emph{ The IWSM and coexisting nodal ring phases.} The Weyl phases are best interpreted in tight-binding model. Other than spin-conserved hopping (denoted by $t_0$ and $t_z$) in $x-y$ plane and $z$ direction, the Raman potential $\mathcal{M}_{x'(y')}$ drives spin-flip hopping along $x'$ ($y'$) direction, with coefficients $t_{\rm so}^{\vec j,\vec j\pm1\hat{e}_{x'}}=it_{\rm so}^{\vec j,\vec j\pm1\hat{e}_{y'}}=\pm i(-1)^{j_{x'}+j_{y'}+j_{z}}t_{{\rm so}}$ of amplitude $t_{{\rm so}}$. The staggered sign $(-1)^{j_{x'}+j_{y'}+j_{z}}$ can be absorbed by a gauge transformation which shifts the Bloch momentum of $|g_{\downarrow}\rangle$ state by $\mathbf{k}_b=(k_0/\sqrt{2},k_0/\sqrt{2},k_0)$~\cite{SI}. The Bloch Hamiltonian reads $\mathcal{H}=\sum_{\bold{q}}\hat{c}^{\dagger}_{\bold{q}}\mathcal{H}(\bold{q})\hat{c}_{\bold{q}},$ where $\hat{c}_{\bold{q}}=(\hat{c}_{\bold{q},\uparrow},\hat{c}_{\bold{q},\downarrow})^T$, $\mathcal{H}(\bold{q})=\bold{h}(\bold{q})\cdot\vec\sigma$, with dimensionless $q_{x',y'}=\pi k_{x',y'}/k'_{0}$, $q_{z}=\pi k_{z}/k_{0}$, and
\begin{eqnarray}\label{tight-1}
\mathcal{H}(\bold{q}) &=&(m_{z}-2t_{0}\cos q_{x'}-2t_{0}\cos q_{y'}-2t_{z}\cos q_{z})\sigma_{z}\nonumber\\
&&+2t_{{\rm so}}\sin q_{x'}\sigma_{x}+2t_{{\rm so}}\sin q_{y'}\sigma_{y}
\end{eqnarray}
describes a set of 2D QAH layers modulated by $q_z$. The Weyl points are located where QAH layers change topology versus $q_z$, given by $\mathcal{H}(\bold{q}_w)=0$, with possible momenta $(q_w^{x'}, q_w^{y'})=(0,0)$, $(0,\pi)$, $(\pi,0)$, or $(\pi,\pi)$ and $q_z=\pm q_w^z$ (or none at all) due to reflection symmetry $z\rightarrow -z$. The number $N$ of Weyl points ranges from $0$ to $8$, 
with the full exact phase diagram for Hamiltonian~\eqref{Hamiltonian1} shown numerically in Fig.~\ref{fig:setup}(d1-d4). {The transition between different Weyl phases is controlled by tuning relative magnitudes of $(m_z,t_0,t_z)$ through $\delta,M,V_0$ and $V_z$.} The IWSM exists in relatively large regions (red-colored areas). For example, for $V_z=2V_0=4E_r$, $M_0/E_r=0.586$, and $m_z/E_r=0.288$, with recoil energy $E_r=(\hbar k_0)^2/2m$, two Weyl points at $\bold{q}_w\simeq(0,0,\pm 0.4\pi)$ are obtained.

{A new novel Weyl phase with coexisting nodal ring and nodal points is predicted when tuning the lattice to be blue-detuned, such that $V_{0,z}<0$ [Zebra-lined area in Fig.~\ref{fig:setup}(d4)]. In this case the relative symmetry and anti-symmetry between the Raman potentials and the lattice are reversed compared with the red-detuned configuration, leading to a key difference in the SO term that is now $\bold h_{\rm so}=(h_x,h_y)=4t_{{\rm so}}\sin q_z( \sin q_{x'},\sin q_{y'})$~\cite{SI}. We then have the Bloch Hamiltonian
\begin{eqnarray}\label{TB_eq}
\mathcal{H}(\bold{q})&=&(m_{z}-2t_{0}\cos q_{x'}-2t_{0}\cos q_{y'}-2t_{z}\cos q_{z})\sigma_{z} \nonumber \\
&&+4t_{{\rm so}}\sin q_z\left( \sin q_{x'}\sigma_{x}+\sin q_{y'}\sigma_{y}\right) 
\end{eqnarray}
With this Hamiltonian we have the similar Weyl points as given by Eq.~\eqref{tight-1}. Interestingly, the nodal ring also emerges at $q_z=0$ ($\pi$) plane and $h_z(\bold{q})|_{q_z=0(\pi)}=0$ for $|m_z-2t_z|<4t_0$ ($|m_z+2t_z|<4t_0$). This gives a new novel Weyl phase with coexisting Weyl points and nodal ring (see Fig.~\ref{SIfig:NLSM}(a1)), as first predicted in this work.

The coexistence of the nodal ring and nodal points is a consequence of the nontrivial relative symmetry between the Raman potentials and the optical lattice. In the blue detuned regime, the Raman potential $M_{x'}$ ($M_{y'}$) is reflection anti-symmetric with respect to each lattice site center along both the $x'$ ($y'$) direction and the $z$ direction. In this way, the spin-flip hopping must jump at least two sites, i.e. one site in $x$-$y$ plane, and one site along $z$ direction. Thus the SO interaction also couples the spin and the quasi-momentum $q_z$. In particular, the relative reflection anti-symmetry ensures that the induced SO coupling must vanish if the quasi-momentum $q_z=0$ and $\pi$, since such two $q_z$ states are reflection symmetric along $z$ direction and must not couple to the spin. This also implies that the nodal ring itself is protected by the reflection symmetry at the $q_z=0$ and $\pi$ planes~\cite{nodalring}, in contrast to the Weyl points which are not symmetry-protected but protected by Chern numbers. On the other hand, the coexisting phase with both nodal ring and Weyl points is characterized by new topological invariants.

\begin{figure*}[t]
\centering
\includegraphics[width=0.8\textwidth]{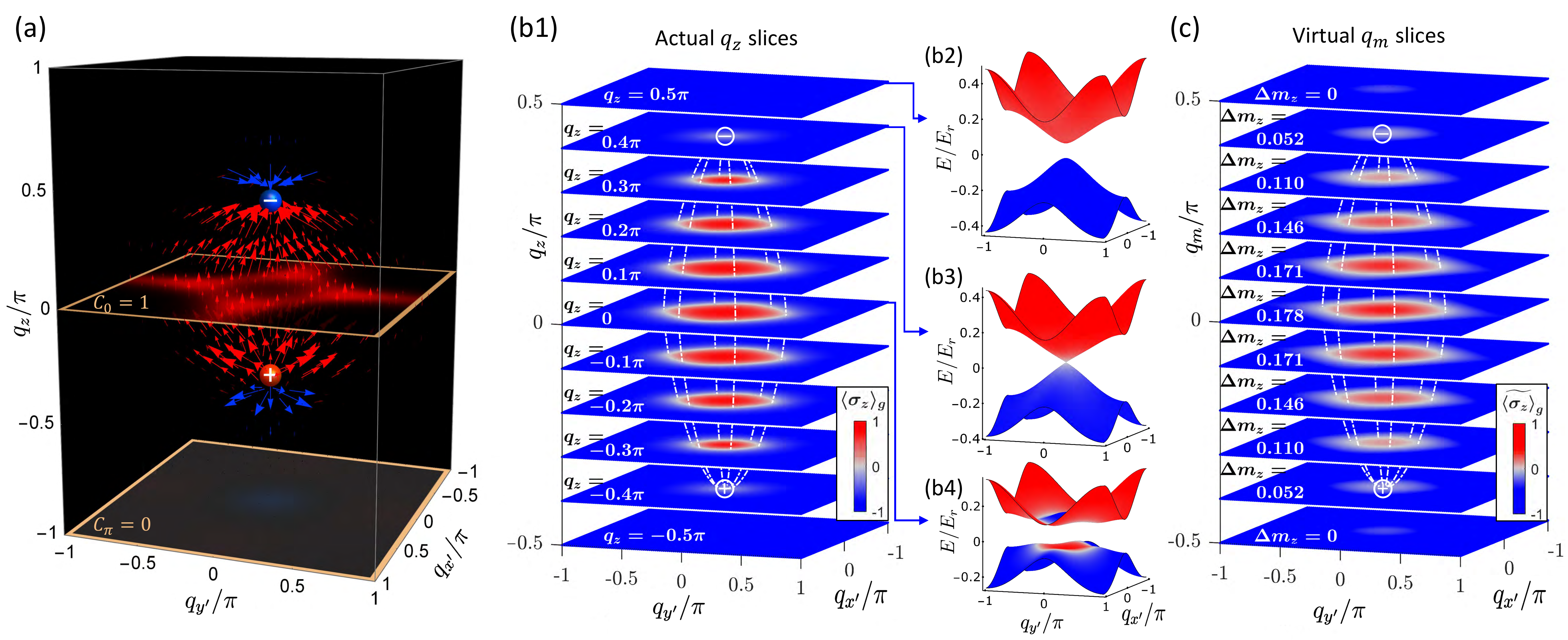}
\caption{\label{fig:Berry_and_spin} Construction of 3D Weyl band topology via {virtual slicing technique.} (a) Weyl points and Berry curvature $\bold{\Omega}(\bold{q})$. Red (blue) spheres located at ${\bold{q}}=(0,0,\pm 0.4 k_0)$ are Weyl points of positive (negative) chirality. Red (blue) arrows and slices correspond to positive (negative) $\Omega_z$. 
(b1) Spin textures of the lower band on different $q_z$ slices. The BIS is read from white rings, where for lower band $\langle{\sigma_z }\rangle_g=0$. (b2)-(b4) 2D Band structures with $q_z=0.5 \pi$, $q_z=0.4 \pi$, and $q_z=0$. 
(c) The $q_z$-integrated 2D spin textures versus $m_z$, as measurable in real experiment. This effectively maps out the 3D Weyl band topology and Weyl points, as compared to (b1). The parameters: $V_0=0.5V_z=2E_r$, $M/E_r=0.586$, and $m_z/E_r=0.288$.}
\end{figure*}
We show that the coexisting phase can be characterized by the nontrivial linking numbers. Note that the 2D bulk of a QAH layer is nontrivial when it has band inversion ring in $q_x$-$q_y$ plane, which is a 1D version of the general notion, band inversion surface (BIS)~\cite{Zhang2018}, defined by $h_z(q_x,q_y)|_{{\rm fix}\ q_z}=0$. The linking number for the coexisting phase can be identified with the following facts. First, the nodal rings at $q_z=0$ and $q_z=\pi$ layers coincide the definition of the band inversion rings for the QAH layers with $q_z=0^+ (0^-)$ and $q_z=\pi^+ (\pi^-)$, respectively (loops $\mathcal{L}_1^\pm$ in Fig.~\ref{SIfig:NLSM}(a2)). For the layers with $q_z\neq0,\pi$, the existence of band inversion ring implies that the layer has nonzero Chern number. Secondly, the Chern number of such 2D QAH layer can reduce to the 1D winding number defined with $\bold h_{\rm so}$ along the 1D band inversion ring~\cite{Zhang2018}. Finally, the existence of Weyl points is directly related to the difference between the Chern numbers at the $q_z=0^+$ (or $0^-$) layer and the $q_z=\pi^+$ (or $\pi^-$) layer. This further gives that the Weyl points can be equivalently characterized through the 1D winding numbers defined on the 1D band inversion rings. In approaching $q_z=0(\pi)$, the winding of $\bold h_{\rm so}$ is replaced with and equals to the winding ($w_1$) of $\tilde{\bold h}_{\rm so}\equiv\partial{\bold h}_{\rm so}/\partial q_z$ along the BIS at $q_z=0(\pi)$ [Fig.~\ref{SIfig:NLSM}(a3)]. Together with the fact that the nodal line can be protected by winding number ($w_0$) of $\bold h(\bold q)$ along a 1D loop [$\mathcal{L}_0$ in Fig.~\ref{SIfig:NLSM}(a1-a2)] enclosing the nodal ring, we reach an important conclusion that the coexisting phase can be characterized by nontrivial linking invariants defined at $q_z=0(\pi)$ plane ${\cal N}_{0(\pi)}=w_0w_1$. 
The number of Weyl points with positive charge (also for negative charge) is then given by
\begin{eqnarray}
N=|w_1(0)-w_1(\pi)|=|{\cal N}_{0}-{\cal N}_{\pi}|
\end{eqnarray}
according to the above analysis. This is also reflected in that a total Berry flux across each $q_z$ plane between the two Weyl points is quantized as $-2\pi$, giving the topological charge $+1$ ($-1$) for the upper (lower) Weyl node [Fig.~\ref{SIfig:NLSM}(b)].
The coexisting Weyl phase is different from the familiar nodal point or nodal line phases~\cite{nodalring}. More details are seen in Supplementary Material~\cite{SI}. 
}

\begin{figure*}[t]
\centering
\includegraphics[width=0.8\textwidth]{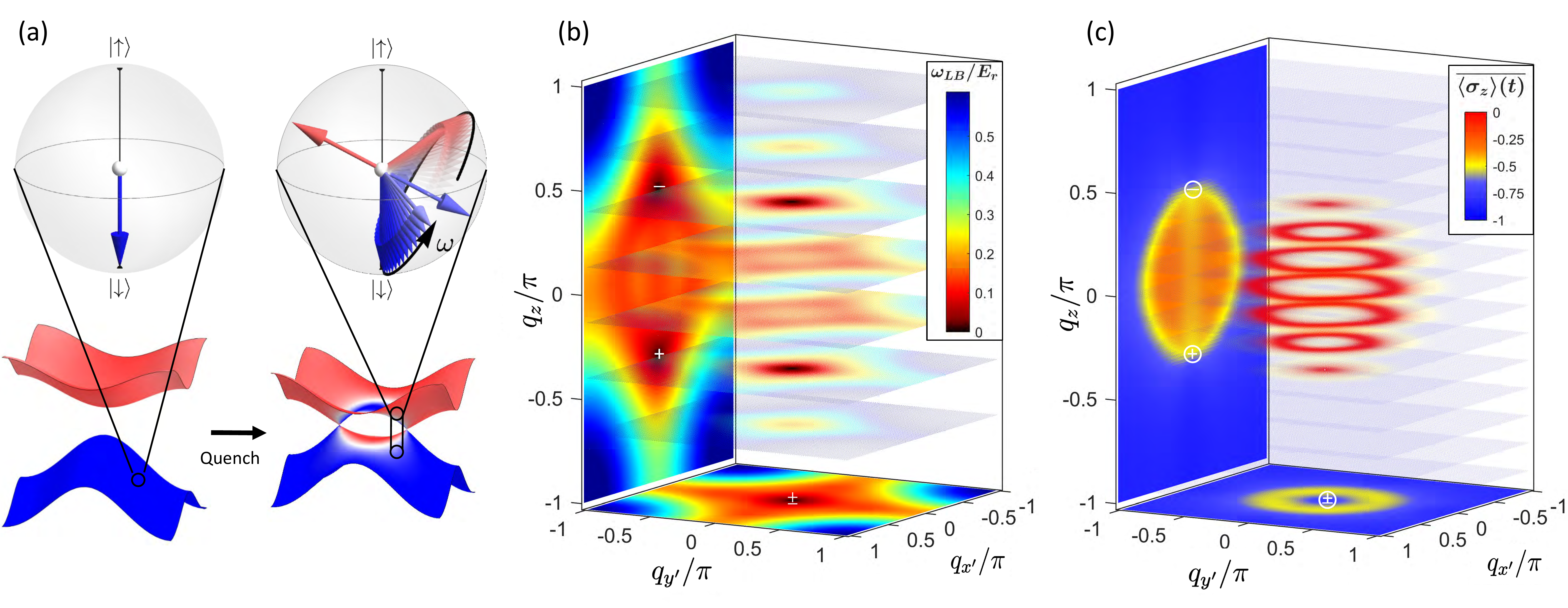}
\caption{\label{fig:quench} Precise measurement of Weyl points by quench dynamics. (a) Pre- and post-quench bands and spin states. The Bloch spheres denote the pre- and post-quench spin states of a certain momentum. The spin precession is depicted by a Larmor-type rotation. (b) Colormap of the spin precession frequency, shown on slices of the 3D BZ. The bottom- and side- walls shows $\omega_{LB}$ in orthogonal projections onto the $q_{x'}-q_{y'}$ plane and $q_{x'}-q_{z}$ plane. Weyl points, being gapless, are captured by two dark spots. (c) Colormap of the averaged spin polarization over certain time. As presented as orthogonal projection view similar to (b), it shows the outline of the BIS as red rings on the slice plot and yellow circles on projection plots.}
\end{figure*}


{\it Measuring 3D Weyl band I: equilibrium scheme.}--We now turn to the detection, and focus on the IWSM phase. We note that the 1D BIS (i.e. band inversion ring) of a QAH phase corresponds to $h_z(q_x,q_y)|_{{\rm fix}\ q_z}=0$ and has vanishing spin polarization $\langle\sigma_z(\bold{q})\rangle|_{\bold{q}\in\rm BIS}=0$ for lower subband~\cite{Zhang2018}. The BISs are resolved in $q_z$-sliced 2D spin textures [Fig.~\ref{fig:Berry_and_spin}(b1)]. When QAH layer changes topology [see Fig.~\ref{fig:Berry_and_spin}(b2-b4)], the BIS shrinks to singular points, rendering Weyl nodes, which are then detected if the configuration of BIS versus $q_z$ can be measured.

We extend our previous virtual slicing technique applied to a $z$-direction continuous system~\cite{Song2019} to the present lattice system. The $q_z$-resolved 2D spin textures in $q_x-q_y$ plane can be effectively detected by measuring the $q_z$-integrated spin textures with varying $m_z$, as compared in Fig.~\ref{fig:Berry_and_spin}(b1) and (c), for which the 3D Weyl band topology and Weyl points are mapped out. First, from the time-of-flight imaging, the $q_z$-integrated 2D spin texture is measured, given by $\widetilde{\langle \sigma_z\rangle}(q_x,q_y) =\frac{1}{2\pi}\int_0^{2\pi}\langle \sigma_z(q_x,q_y) \rangle \text{d}q_z$.
For the layer at $q_z=\pi/2$, the band inversion ring is determined by $m_z=2t_0(\cos q^{B}_{x'}+\cos q^{B}_{y'})$. Then, for $q_z=\pi/2+\delta q_z$ and $(q_{x'},q_{y'})=(q^{B}_{x'},q^{B}_{y'})$ the effective Hamiltonian reads ${\cal H}_{\rm eff}(\bold{q})=2t_z\sin{\delta q_z}\sigma_z+2t_{\rm so}\sin{q_{x'}}\sigma_x+2t_{\rm so}\sin{q_{y'}}\sigma_y$, which has an emergent magnetic group symmetry ($\mathcal{M}$) defined by
\begin{eqnarray}
\mathcal{M}^{-1}{\cal H}_{\rm eff}(\delta q_z)\mathcal{M}={\cal H}_{\rm eff}(-\delta q_z), \ \mathcal{M}=\sigma_x K,
\end{eqnarray}
with $K$ the complex conjugate. Due to the symmetry the two Bloch states $|u_{\delta q_z}(q_{x'},q_{y'})\rangle$ and $|u_{-\delta q_z}(q_{x'},q_{y'})\rangle$ have the same energy but opposite spin-polarizations along $z$ direction. Thus the $q_z$-integrated spin-polarization renders that of the $\mathcal{M}$-symmetric 2D layer at $q_z=\pi/2$, i.e. $\widetilde{\langle \sigma_z \rangle}(q_{x'},q_{y'}) =\langle \sigma_z(q_{x'},q_{y'})\rangle|_{q_z=\pi/2}$.
Finally, tuning $m_z$ is equivalent to scanning $q_z$ as $\langle \sigma_z \rangle_g|_{\delta q_z}^{m_z}=\langle \sigma_z \rangle_g|_{\delta q_z=0}^{m_z-\Delta m_z}$, with $\Delta m_z = -2t_z \sin \delta q_z$. This equals to scanning a virtual momentum $q_m=\pm\frac{k_0}{\pi}\arccos{\left(\frac{\Delta m_z}{2t_z}\right)}$. The spin-polarization changes sign across BIS, and satisfies
\begin{eqnarray}
{\rm sgn}\left(\langle \sigma_z \rangle|_{\delta q_z}^{m_z}\right)={\rm sgn}\left(\widetilde{\langle \sigma_z \rangle}|_{\delta q_z=0}^{m_z-\Delta m_z}\right).
\end{eqnarray}
Therefore, the virtual slices of the fusiform BIS shown in Fig.~\ref{fig:Berry_and_spin}(c) by varying $m_z$ effectively map out the Weyl band topology and Weyl points. {The topological charge of Weyl point equals the Chern number change of QAH layers across the Weyl node in $q_z$ direction.}

{\it Measuring 3D Weyl band II: dynamical scheme.}--We further propose to detect the Weyl points by quench dynamics.
The system is initialized in the fully polarized trivial regime by a large $m_z$, with atoms occupying $|g_{\downarrow}\rangle$ state, and then suddenly tuned to the target IWSM phase with small $m_z$ [Fig.~\ref{fig:quench}(a)]. The quench dynamics at $\bold{q}$ reads $\langle\sigma_z(\bold{q})\rangle(t)=\langle g_{\downarrow}e^{i\mathcal{H}(\bold{q})t/\hbar}|\sigma_z|e^{-i\mathcal{H}(\bold{q})t/\hbar}g_{\downarrow}\rangle$. For the short-term dynamics, with damping effects being negligible, the spin precession is obtained by~\cite{SI}
\begin{eqnarray}
\langle\sigma_z(\bold{q})\rangle(t)=-\langle\sigma_z(\bold{q}\rangle_{g}^2-\left(1-\langle\sigma_z(\bold{q}\rangle_{g}^2\right)\cos \omega(\bold{q}) t.
\end{eqnarray}
Here $\langle\sigma_z\rangle_{g}$ is the spin-polarization of ground state of the post-quench Hamiltonian and the frequency $\omega(\bold{q})$ equals the local band gap. 
The Weyl points with $\omega(\bold{q}_w)=0$ can be determined by projection measurements. First, we project $\langle\sigma_z\rangle(t)$ in the $q_{x'}$-$q_{y'}$ plane, and measure the minimal frequency $\omega_{LB}={\min}[\omega(q_z)]$ with every fixed $(q_{x'},q_{y'})$, which determines the Weyl momentum $(q_{x'}^w,q_{y'}^w)$ in $q_{x'}$-$q_{y'}$ plane. Further, we project the measurement in $q_{x'}$-$q_{z}$ plane and determine $(q_{x'}^w,q_{z}^w)$ from the minimal frequency $\omega_{LB}={\min}[\omega(q_{y'})]$ at every fixed $(q_{x'},q_{z})$. The two steps give the Weyl momentum $\bold{q}^w$, as illustrated in Fig.~\ref{fig:quench}(b).

The BIS can also be dynamically measured by spin dynamics. The time-averaged spin polarization $\overline{\langle\sigma_z\rangle}(t)=\frac{1}{t}\int_{0}^{t}\langle\sigma_z\rangle(\tau)\mathrm{d}\tau=-\langle\sigma_z\rangle_{g}^2$ is negative and approaches zero only on BIS~\cite{Zhang2018}. Similar, one cannot directly resolve the BIS dynamically in 3D momentum space, but the projection measurement onto 2D spaces can be achieved, with the third dimension being integrated out. Projection to two orthogonal planes, e.g. $q_{x'}$-$q_{y'}$ and $q_{x'}$-$q_{z}$ planes can extract all Weyl nodes in the bulk, as shown in Fig.~\ref{fig:quench} (c).

Before conclusion we make a comparison between the equilibrium and dynamical schemes for the detection. The equilibrium scheme provides a more straightforward and intuitive detection by reconstructing the 3D band topology. However, the dynamical scheme usually can have a higher precision in detecting the Weyl points. The reason is two-fold. First, in the quench study, the initial state is prepared in a deep trivial regime, which can be ideally realized by a large Zeeman splitting, not affected by the perturbations due to unperfect conditions like thermal effects and noise. Secondly, the short-term quench dynamics after suddenly changing parameters should be approximately unitary, which enables a nearly ideal measurement in experiment. In comparison, for the equilibrium scheme, the initial state should be prepared in the topological regime, which has vanishing gap around the Weyl points, and can be easily affected by the unperfect conditions. Such unperfect conditions also reduce the resolution of the measurements. With the advantages in both the preparation of initial state and the measurement, similar to the previous study of a QAH insulator~\cite{Zhang2018,Sun2018b}, the dynamical scheme is expected to have a higher precision in the detection. More details of the dynamical scheme of detection are presented in Supplementary Material~\cite{SI}.

{\it Conclusion.}---We have proposed to realize and detect 3D SO coupling and IWSM based on ultracold atoms, which were not achieved previously. A new Weyl phase with coexisting Weyl nodal points and nodal ring is also uncovered and shown to be characterized by nontrivial linking numbers. This coexisting phase is different from the known 3D semimetals, and may host the new exotic Weyl physics and deserves future study. The study presented in this work has led to the first experimental realization of the IWSM band with minimal Weyl points and the first experimental realization of 3D SO coupling for ultracold atoms~\cite{note}. Realization of the IWSM and 3D SO coupling shall open up a broad avenue to explore unambiguously the novel Weyl physics, and for example, may be helpful to resolve the debating issue of the dissipationless chiral magnetic effect~\cite{Fukushima2008,Franz2013} which was widely studied in condensed matter physics~\cite{CME1,CME2,CME3,CME4,CME5,CME6} (a concrete study is presented in Supplementary Material~\cite{SI}), and further interacting phases in this optimal platform.

This work was supported by National Natural Science Foundation of China (11825401, 11761161003, and 11921005), National Key R\&D Program of China (2016YFA0301604), and Strategic Priority Research Program of CAS (Grant No. XDB28000000).

\newpage
\onecolumngrid

\renewcommand{\thesection}{S-\arabic{section}}
\setcounter{section}{0}  
\renewcommand{\theequation}{S\arabic{equation}}
\setcounter{equation}{0}  
\renewcommand{\thefigure}{S\arabic{figure}}
\setcounter{figure}{0}  

\indent

In this supplementary material we provide the details of deriving the effective Hamiltonian, which is valid for both fermions and bosons (e.g. $^{87}$Rb atoms), the effect of reflection-symmetry breaking, the coexisting nodal point and nodal ring semimetal, the techniques for measurement, and the chiral magnetic effect.

\addtocontents{toc}{\protect\setcounter{tocdepth}{2}}
\section{I. The Model of Realization}
\label{SI:setup}

\subsection{A. Laser Setup}
\label{SI:lasers}

As shown in  Fig.1(a) in the main text, two beams forms orthogonal standing wave in the $x-y$ plane, while another beam forms standing wave in the $z$  dimension.
The frequencies of the beams are $\omega_1$ for the light $\bold{E}_{xy}$, $\omega_2=\omega_1+\delta \omega$ for the beam $\bold{E}_{xz}$, and $\omega_3=\omega_1+\Delta\omega$ for $\bold{E}_z$ in the $z$ direction, where $\delta\omega$ matches the Zeeman splitting of the two spin states,
$\mu_B g_F B/\hbar$, by a small two-photon detuning $\delta$, and  $\Delta\omega$ is in the magnitude of a few MHz. The polarization of the beams are as shown in the Fig.1(a) in the main text.
We can conveniently write the light fields as
\begin{align}
    \bold{E}_{xz} &= (\bold{\hat{z}}E_{xz} e^{{\rm i}k_{0}x}
    +\bold{\hat{z}}E_{yz} e^{-ik_{0}y}
    +\bold{\hat{z}}E_{yz} e^{ik_{0}y}
    + \bold{\hat{z}}E_{xz} e^{-ik_{0}x})e^{-i\omega_3 t}  \\ \nonumber
    &= E_{0}\bold{\hat{z}}(\cos k_{0}x+\cos k_{0}y) e^{-i\omega_3 t}, \\ \nonumber
    \bold{E}_{xy} &= (-i \bold{\hat{y}}E_{xy} e^{{\rm i}k_{0}x}-i \bold{\hat{x}}E_{yx} e^{-ik_{0}y} +i \bold{\hat{x}}E_{yx} e^{ik_{0}y}+ i \bold{\hat{y}}E_{xy} e^{-ik_{0}x})e^{-i\omega_1 t} \\ \nonumber
    &= E_{0} (\bold{\hat{y}}\sin k_{0}x - \bold{\hat{x}}\sin k_{0}y) e^{-i\omega_1 t},\\ \nonumber
    \bold{E}_{z}&= \left[ e^{ i(k_{0}z-\omega_2 t)} + e^{ i(-k_{0}z-\omega_2 t)} \right] (E_{zx} \bold{\hat{x}}+ E_{zy} i \bold{\hat{y}}) \\ \nonumber
    &= E_{z}(\bold{\hat{x}}+i\bold{\hat{y}})\cos k_{0}z e^{-i\omega_2 t},
\end{align}
where  $E_0 = 2E_{xy}=2E_{yx}=2E_{xz}=2E_{yz}$ and $E_z =2E_{zx}=2E_{zy}$ . Here we use $E_{\alpha \beta}$ to represent the laser propagating along $\alpha$-direction with $\beta$-polarization.  The wavenumbers are all denoted as $k_0$ because the frequency difference is comparatively small that no observable phase difference is accumulated over the propagation length of the laboratory light path. We further denote $\bold{E}_{x-y}=\bold{E}_{xz}+\bold{E}_{xy}$ for convenience.

\subsection{B. Lattice Potential and Raman Field}

The spin-independent optical potential  for typical detuning
$\Delta$ is proportional to light intensity,
$\mathcal{V}_{latt} \propto 
(\bold{E}_{x-y}^*\bold{E}_{x-y}+\bold{E}_z^*\bold{E}_z)/\Delta $. For detailed calculation, we sum up coupling induced by the $\sigma^+ / \pi / \sigma^-$  components.
Note that the external magnetic field points in $+\bold{x}$ direction, so the $\sigma^+ / \pi / \sigma^-$ components have polarization of $\frac{\hat{y}+i\hat{z}}{2} / \hat{x} / \frac{\hat{y}-i\hat{z}}{2}$, respectively.  The lattice potentials for $|g_{\uparrow}\rangle$ and $|g_{\downarrow}\rangle$ are deduced as follow
\begin{align}
V_{\downarrow} &=  \sum_{j=\frac{1}{2},\frac{3}{2}} \sum_F \frac{1}{\Delta_{j}} \left( |\Omega_{\downarrow F,xy}^{(j)}|^2 + |\Omega_{\downarrow F,yx}^{(j)}|^2 + |\Omega_{\downarrow F,xz}^{(j)}|^2 + |\Omega_{\downarrow F,yz}^{(j)}|^2 +
|\Omega_{\downarrow F,z}^{(j)}|^2\right)  \\   \nonumber
&= \frac{2}{3} (\frac{\alpha_{D_2}^2}{\Delta_{3/2}} + \frac{\alpha_{D_1}^2}{\Delta_{1/2}} ) \left(\cos k_0x \cos k_0y E_0^2 + \cos^2 k_0z E_z^2 \right)  \\ \nonumber
&= -V_0 \cos k_0x \cos k_0y - V_z \cos^2 k_0z,
\end{align}
where $\alpha_{D_2}^2 \widetilde{=}\, 2\alpha_{D_1}^2$ are transition dipole matrix elements can be found in ~\cite{steck2001rubidium}. Similarly, for $|g_{\uparrow}\rangle$ atoms we also have $V_\uparrow =V_\downarrow$. The key point here is that our lattice potential is deformed and aligned in the diagonal direction. Only in this deformed lattice, the following  Raman fields have the symmetry we need to realize Weyl semimetal.

The Raman potentials $\mathcal{M}(\bold{r})$ are generated by a $\Lambda$-type scheme from bichromatic lights $\bold{E}_{xy}$ and $\bold{E}_z$ ($\bold{E}_{xz}$ does not participate in Raman transitions between the ground states) that couples $|g_{\uparrow}\rangle$  to $|g_{\downarrow}\rangle$, as shown in Fig. 1(b). It can be explained as an effective Zeeman field.
For $^{87}$Rb atoms, the effective Zeeman field has the following form
\begin{align}
\mathcal{M}(\bold{r}) &= \sum_{j=\frac{1}{2},\frac{3}{2}} \sum_{F} \left(\frac{\Omega_{\downarrow F,zy}^{(j)*}\Omega_{\uparrow F,yx}^{(j)}}{\Delta_{j}} + \frac{\Omega_{\downarrow F,zx}^{(j)}\Omega_{\uparrow F,xy}^{(j)}}{\Delta_{j}}\right) =\left(\frac{\alpha_{D_{2}}^{2}}{12}\frac{1}{\Delta_{3/2}}-\frac{\alpha_{D_{1}}^{2}}{6}\frac{1}{\Delta_{1/2}}\right)\left(\frac{E_{zy}^{*}}{\sqrt{2}}E_{yx}+E_{zx}^{*}\frac{E_{xy}}{\sqrt{2}}\right)\\ \nonumber
&= M_{0}\cos k_{0}z(\sin k_{0}x\sigma_{x}+\sin k_{0}y\sigma_{y}).
\end{align}

\subsection{C. Weyl Hamiltonian}
\label{SI:Ham}
Combining the results of Raman field $\mathcal{M}(\bold{r})$ and lattice potential $\mathcal{V}_{latt}(\bold{r})$, we arrive at the target Hamiltonian
\begin{align}
\mathcal{H}&=\left[\frac{\bold{p}^2}{2m}+\mathcal{V}_{latt}(\bold{r})\right]\mathbb{1} +\mathcal{M}_x(\bold{r}) \sigma_x+\mathcal{M}_y(\bold{r}) \sigma_y +m_z\sigma_z.
\end{align}
It is apparent that the lattice potential in the $x-y$ plane is diagonally aligned.  To align the lattice with the coordinate axes, we rotate the axis in the $x-y$ plane by $\pi/4$ such that $x'=(x-y)/\sqrt{2}$ and $y'= (x+y)/\sqrt{2}$.
Under this rotation in the real space, the Hamiltonian can be written in the square lattice as
\begin{align}
\mathcal{H}&=\left(\frac{\bold{p}^2}{2m} -V_0 \cos^2k'_{0}x'- V_0 \cos^2k'_{0}y' - V_z \cos^2 k_0z\right)\mathbb{1}+m_z\sigma_z \notag \\
&{} \quad +\sqrt{2} M_{0}\cos k_{0}z \left[\sin (k'_{0}x')\cos (k'_{0}y') \sigma_x+\cos(k'_{0}x')\sin (k'_{0}y') \sigma_y \right].
\end{align}
Here we have $k'_0 = k_0 /\sqrt{2}$ after rotation.

To get a more intuitive understanding of our Hamiltonian, we take it to the tight-binding(TB) picture. Under the deep lattice limit, the atoms are positioned at the bottom of the lattice sites, while the spin-reserved (spin-flipping) hopping terms are decided by the inter-site overlap of Wannier wavefunctions (plus Raman potential), respectively.
We observe that the Raman potentials are in staggered superposition over the lattice cells. Therefore,
a staggered gauge transformation $U = e^{{i}(k'_0x' + k'_0 y' + k_0z)|g_{\downarrow}\rangle\langle g_{\downarrow}|}$ can be applied on the spin-down state as was performed in 2D SOC system~\cite{Liu:2014hj}.
Then the tight-binding Hamiltonian in the momentum space can be written as
\begin{align}
\label{TB_eq}
H(\bold{q}) & =\bold{h}(\bold{q})\cdot \bold{\sigma} \nonumber \\
&= (m_{z}-2t_{0}\cos q_{x'}-2t_{0}\cos q_{y'}-2t_{z}\cos q_{z})\sigma_{z} +2t_{{\rm SO}}\sin q_{x'}\sigma_{x}+2t_{{\rm SO}}\sin q_{y'}\sigma_{y}.
\end{align}
The band-inversion-surface(BIS) satisfies the equation $h_z(\bold{q})=0$.
We can see that the Weyl points should be located at the intersections of the BIS and one of these four lines along the $q_z$ direction: $(q_{x'}, q_{y'})=(0,0)$, $(0,\pi)$, $(\pi,0)$, or $(\pi,\pi)$. On each line there must be a pair of Weyl points of opposite $q_z$ and opposite Weyl charge (or none at all), due to reflection symmetry of the BIS.

Here we are especially interested in the system that have only two Weyl points, which can be realized in the conditions $|m_z| >2t_z$ or
$-2t_0<4t_0 -|m_z|<2t_z$. As an example, under these chosen parameters $V_0=2E_r$, $V_z=4E_r$, $M=0.586E_r$, we can tune $m_z=0.288 E_r$ so that $N=2$, the only two Weyl points are located at $\bold{q}=(0,0,\pm 0.4\pi)$. On the other hand, there are two types of $N=0$ phases. The first type is trivial and corresponds to the regime that $|m_z|>4|t_0|+2|t_z|$ [white areas in the Fig. 1(d) of main text]. In this case, any 2D layer with fixed $q_z$ has zero Chern number. The second type is a 3D Chern insulator and corresponds to the regime that $0<|m_z\pm2t_z|<4|t_0|$ [the black area in Fig. 1(d) of main text]. In this case, each 2D layer with fixed $q_z$ has the same nonzero Chern number.

\subsection{D. Weyl phase with coexisting Weyl points and nodal ring}
\label{SI:Nodal}


For overall blue detuned cases, where $(\frac{2}{\Delta_{3/2}} + \frac{1}{\Delta_{1/2}} )$ is positive, $V_0$ and $V_z$ becomes negative so that the peaks and valleys of the potential are switched. However, the Raman field remains the same. As the result, the Raman-potential-induced spin-flipped hopping vanishes for the in-plane nearest neighbor couplings, but instead appears on the second-nearest neighbor sites respectively in the adjacent planes along $z$ axis. Then following the same process of deriving the Bloch Hamiltonian in the red-detuned configuration, we arrive at the Bloch Hamiltonian as
\begin{align}
H(\bold{q}) = \bold{h}(\bold{q})\cdot\bold{\sigma}= (m_{z}-2t_{0}\cos q_{x'}-2t_{0}\cos q_{y'}-2t_{z}\cos q_{z})\sigma_{z} +4t_{{\rm SO}}\sin q_z\left( \sin q_{x'}\sigma_{x}+\sin q_{y'}\sigma_{y}\right).
\end{align}
It is readily seen that the BIS remains the same as the red-detuned case, and the Weyl points, when existing, are still located at the intersection points between the BISs and $(q_{x'}, q_{y'})=(0,0)$, $(0,\pi)$, $(\pi,0)$, and $(\pi,\pi)$ lines. Moreover, nodal ring also emerges at $q_z=0$ ($\pi$) plane and $h_z(\bold{q})|_{q_z=0(\pi)}=0$ for $|m_z-2t_z|<4t_0$ ($|m_z+2t_z|<4t_0$). This gives a new novel Weyl phase with coexisting Weyl points and nodal ring. As we study in the main text, the coexisting phase can be characterized by the nontrivial linking number defined at $q_z=0(\pi)$ plane ${\cal N}_{0(\pi)}=w_0w_1$, where $w_{1(0)}$ is the winding number of topological spin texture along 1D loop along nodal ring (1D loop enclosing nodal ring). We show this result in more detail below.

The first topological number $w_0$ is the winding number of an infinitesimal circle $\mathcal{L}_0$ that threads the nodal line, illustrated as the dark-red circle in Fig.~2(a) of main text. The linear expansion of the Hamiltonian around the center gapless point renders a result of $H\left(\Delta\bold{q}\right)=\frac{\partial \bold{h}(\bold{q})}{\partial \bold{q}}\cdot \Delta\bold{q} $ and thus the loop integral on $\mathcal{L}_0$ will always trace out a great circle on the Bloch sphere, which in turn gives a Berry phase of $\pm \pi$. As usual, we obtain the following $\text{Z}_2$-invariant:
$$w_0 = \frac{1}{\pi}\oint_{\mathcal{L}_0}\bold{A}(\bold{q})\text{d}\bold{q} = 1,
$$
where $\bold{A}(\bold{q})$ is the Berry connection of the occupied band:
$$\bold{A}(\bold{q}) = -i \langle{\psi_1(\bold{q})}|{\partial \bold{q}}|{\psi_1(\bold{q})}\rangle$$

For the second topological invariant $w_1$, we consider firt the winding number on a loop on BIS that is closely parallel to the nodal loop, illustrated as $\mathcal{L}_1^{\pm}$ in Fig.~2(a2) of main text. Note that we cannot take the integral directly on the nodal line (because the spin texture vanishes), but we can take the nodal line as the limit, either from the $q_z\rightarrow 0^+$ or $q_z\rightarrow 0^-$, which will give us the same Berry phase from the loop integral:
$$\gamma_1 = \oint_{\mathcal{L}_1^\pm}\bold{A}(\bold{q})\text{d}\bold{q}=\lim_{\delta q_z\rightarrow 0}\oint_{\mathcal{L}_1}\bold{A}(\bold{q+\delta q_z \hat{\bold{z}}})\text{d}\bold{q}$$
The spin-texture along the nodal line $\mathcal{L}_1$ is confined within the $q_z = 0$ plane, thus we can throw away the limit by introduce the following effective dynamical Berry phase:
$$\gamma_1 = \oint_{\mathcal{L}_1}\tilde{\bold{A}}(\bold{q})\text{d}\bold{q}=-\pi,$$
Where $\tilde{\bold{A}}(\bold{q})$  is the Berry connection of the occupied band to the following effective Hamiltonian:
$$\tilde{H}(q_{x'},q_{y'}) = \left. \frac{\partial H(\bold{q})}{\partial q_z}\right|_{q_z=0}= 4t_{{\rm SO}}\left( \sin q_{x'}\sigma_{x}+\sin q_{y'}\sigma_{y}\right).$$
The corresponding effective spin texture is given by the vector field $\tilde{\bold{h}}(q_{x'},q_{y'}) = \partial \bold{h}(\bold{q})/\partial q_z|_{q_z=0}$
along the nodal line. Thus the winding number $w_1 = \gamma_1/\pi = -1$, which is a $\text{Z}$-invariant. According to the above analysis the number of Weyl points with positive charge (also for negative charge) is then given by
\begin{eqnarray}
N=|w_1(0)-w_1(\pi)|=|{\cal N}_{0}-{\cal N}_{\pi}|.
\end{eqnarray}

We provide further the physical interpretations of the two winding numbers and show that $w_0$ is indeed a $\text{Z}_2$-invariant for the present two band model that is protected by reflection symmetry, and $w_1$ is a $\text{Z}$-invariant. Our Hamiltonian is invariant under reflection with respect to the $z=0$ plane, $\mathcal{M}_z=i\sigma_z$, which acts on the spin space as $\sigma_z$. It is easy to verify the following: $\sigma_z H(\bold{q}) \sigma_z^{-1} = H(-\bold{q})$, giving the two reflection-invariant planes at the $q_z = 0$ and $q_z = \pi$. The energy eigenstates at the symmetric planes are also $\mathcal{M}_z$ eigenstates of quantum number $\pm 1$ ($+1$ for even-reflection-symmetric, or equivalently, spin-up states), and thus do not hybridize. We note that $w_0 = N_{in}-N_{out}$, where $N_{in}$($N_{out}$) is the number of occupied  even-reflection-symmetric bands inside(outside) the nodal loop~\cite{Fang_2016}. This tells us that our nodal line is stable under perturbation that preserves $\mathcal{M}_z$ reflection symmetry, but may break into Weyl node pairs when other forms of perturbation is introduced.

For $w_1$, we consider the total Berry flux $\Phi_B(q_z)$ penetrating the $q_z = \text{const}$ plane. If we do a surface integral of Berry curvature (as shown in Fig.~2(b)) of main text, we may readily find that $\Phi_B(q_z) = \int\int \Omega_z(q_{x'},q_{y'},q_{z})dq_{x'}dq_{y'}=-2\pi$ for any $0<|q_z|<\pi/2$. For the $q_z = 0$ plane, at first sight $\Phi_B(0)=0$ because Berry curvature vanishes everywhere on the $q_z=0$ plane except for on the nodal line, which has a surface area of zero. However, since the total Berry's charge of the nodal loop is zero, the Berry curvature field should have no divergence near $q_z=0$. We should have then
$$\Phi_B(0)=\lim_{\epsilon\rightarrow 0}\Phi_B(\epsilon) = -2\pi.$$
Apparently, the latter is correct due to the fact that the Berry curvature field narrows down and squeezes through the nodal line, diverges in magnitude right on the nodal ring but preserves the total flux. As a result, the surface integral of Berry curvature is no longer a good definition for the total Berry flux on $q_z = 0$. To calculate the Berry flux going through  the nodal loop, we apply the Stokes theorem: $$\Phi_B(0)=  \oint_{\mathcal{L}_3}\bold{A}(\bold{q})\text{d}\bold{q} - \oint_{\mathcal{L}_2}\bold{A}(\bold{q})\text{d}\bold{q},$$
where $\mathcal{L}_2$ ($\mathcal{L}_3$) are two paths within the $q_z=0$ plane that goes slightly inside (outside) the nodal loop. There is a subtle gauge choice here: if we mark the south (north) pole of the Bloch sphere as the singularity point, then the Berry connection $\bold{A}(\bold{q})$ outside (inside) the nodal loop would be ill-defined, but if we mark the singularity anywhere but the poles then there will be a branch-cut that links the Weyl nodes onto the nodal loop, which introduces extra flux. To avoid this problem, we split the difference into two parts and apply different gauge choice accordingly.
\begin{align*}
   \Phi_B(0) =& \left[\oint_{\mathcal{L}_3}\bold{A}(\bold{q})\text{d}\bold{q} - \oint_{\mathcal{L}_1^+}\bold{A}(\bold{q})\text{d}\bold{q}\right] + \left[\oint_{-\mathcal{L}_1^+}\bold{A}(\bold{q})\text{d}\bold{q} - \oint_{-\mathcal{L}_2}\bold{A}(\bold{q})\text{d}\bold{q}\right] \\
   =& \,\, 2 \oint_{\mathcal{L}_1^+}\bold{A}(\bold{q})\text{d}\bold{q} = 2\gamma_1 = 2\pi w_1 = -2\pi
\end{align*}
Where in the first(second) bracket, we used the gauge where the singularity is at the north(south) pole, and in the second line the singularity is at the south pole. If we compare this with the definition of Chern number on a $q_z = \text{const}$ slice, which is $\mathcal{C}(q_z)=\frac{1}{2\pi}\Phi_B(q_z)$, we get:
$$w_1 = \lim_{\epsilon\rightarrow 0}\mathcal{C}(\epsilon)$$

This reconfirms the aforementioned result that $w_1$ is equivalent to the Chern number of an adjacent plane, which acts as a bridge that connects the Weyl point topology with the winding along the nodal ring.
We know that the intersection of the BIS and either the $q_z=0$ or the $q_z=\pi$ plane renders a nodal line; and the $w_1$ of each nodal line marks the Berry flux going through the plane (if there is no nodal ring in a certain plane means that the net Berry flux of that plane is zero). 
It can be seen that $w_0$ and $w_1$ forms a nontrivial linking number. The number of Weyl points with positive charge (also for negative charge) is given by $N=|w_1(0)-w_1(\pi)|=|{\cal N}_{0}-{\cal N}_{\pi}|$. The coexisting Weyl points and nodal ring requires that ${\cal N}_{0}$ and ${\cal N}_{\pi}$ are different.

\subsection{E. Reflection symmetry breaking and type-II Weyl semimetal}
\label{SI:type-2}

The reflection symmetry of the Weyl semimetal can be easily broken. For this we slightly modify the optical Raman lattice. We consider a standing-wave beam $\bold{E}_{zV}$, of wave-number $k_V$, to form the 1D lattice in the $z$ direction. This beam does not contribute to the Raman coupling by setting that it has a large frequency difference from all other beams. In particular, the beam $\bold{E}_{zV}$ can be far detuned and does not lead to heating in the system. Further, we apply another running-wave beam $\bold{E}_{zR}$ along $z$ direction, which together with the $x$-$y$ lights (wave-number $k_0$) generates the two Raman potentials. In this way,
the Raman potentials take the form$\mathcal{M}(\bold{r})=M_{0} e^{k_{0}z}(\sin k_{0}x\sigma_{x}+\sin k_{0}y\sigma_{y})$,
where $M_0 \propto E_{zR} $ and $m_z=\delta/2$. After applying $\pi/4$ rotation, we have
\begin{align*}
H&=\left(\frac{\bold{p}^2}{2m} -V_0 \cos^2k_{0}u- V_0 \cos^2k_{0}v - V_z \cos^2 k_V z\right)\mathbb{1}+m_z\sigma_z\\
&+\sqrt{2} M_{0} e^{k_{0}z} \bigr[\sin (k_{0}u)\cos (k_{0}v) \sigma_u+\cos(k_{0}u)\sin (k_{0}v) \sigma_v\bigr].
\end{align*}
In the $x$-$y$ plane the Hamiltonian has the same form as the case with reflection symmetry, but along $z$-direction we observe that a phase difference of $e^{i\pi k_0/k_V}$ between each $z$ layer appears in the tight-binding model. To remove the phase factor $e^{i \pi j_z k_0/k_V}$ in the spin-flip hopping terms, we further apply a translation $U=\exp \left[  i\pi j_{z}(1+k_{0}/k_V \sigma_z)/2 \right]$ to the basis. Then the tight-binding Hamiltonian in the momentum space becomes
\begin{eqnarray}
\mathcal{H}(\bold{q})&=& \left(m_{z}-2t_{0}\cos q_{u}-2t_{0}\cos q_{v}-2t_{z}\sin\frac{\phi_{z}}{2}\cos q_{z}\right)\sigma_{z} \nonumber\\
&& + 2t_{{\rm SO}}\sin q_{u}\sigma_{x}+2t_{{\rm SO}}\sin q_{v}\sigma_{y} -2t_{z}\cos\frac{\phi_{z}}{2}\sin q_{z} \otimes \mathbb{1},
\end{eqnarray}
where $\phi_{z}=k_{0}\lambda_V/2=\pi k_{0}/k_{V}$. The last term in the above Hamiltonian breaks the reflection symmetry and can lead to energy imbalance between Weyl points with different quasimomentum $q_z^w$.

We consider the case with two Weyl point, which is obtained when $|m_z|>2t_z\sin\frac{\phi_{z}}{2}$, and
$|m_z- 4t_0|<2t_z\sin\frac{\phi_{z}}{2}$. Expanding the base term $\bold{h}_{base}(\bold{q})=-2t_{z}\cos\frac{\phi_{z}}{2}\sin q_{z} $ around the
Weyl point $\left(0,0,\pm \arccos{\frac{m_z- 4t_0}{2t_z\sin\frac{\phi_{z}}{2}}}\right)$, we would obtain a non-zero gradient term $\mp 2t_z \cos\frac{\phi_{z}}{2} \sqrt{1-\left({\frac{m_z- 4t_0}{2t_z\sin\frac{\phi_{z}}{2}}}\right)^2}$,
which means that the degree to which the two weyl cones are tilted can be easily controlled by tuning $m_z$. When $m_z$ is tuned gradually from $4t_0+2t_z\sin\frac{\phi_{z}}{2}$ down to $4t_0-2t_z\sin\frac{\phi_{z}}{2}$, the Weyl cones undergo a Lifshitz transition in the manner of $\text{type-II} \rightarrow \text{type-I} \rightarrow \text{type-II}$.


\section{II. Measuring the 3D Weyl band: Virtual Slicing Technique}
\label{SI:ToF}

There is an equivalent method to scan the $q_z$  for BIS measurement: the \emph{Virtual Slicing Technique}. While a 3D ToF imaging is technically challenging, it is easy to perform a 2D  ToF  imaging of  spin polarization with $q_z$ projected out: $\widetilde{\langle \sigma_z \rangle}=\frac{1}{2\pi}\int_0^{2\pi}\langle \sigma_z \rangle \text{d}q_z$. We prove that scanning $q_z$ in $\langle \sigma_z \rangle_g$  can be mimicked by tuning $m_z$ in $\widetilde{\langle \sigma_z \rangle}_g$. First, ${\rm sgn}(\langle \sigma_z \rangle_g)={\rm sgn}(h_z(\bold{q}))={\rm sgn}(m_{z}-2t_{0}\cos q_{x'}-2t_{0}\cos q_{y'}-2t_{z}\cos q_{z})$. Therefore, ${\rm sgn} (\widetilde{\langle \sigma_z \rangle}_g )={\rm sgn}(m_{z}-2t_{0}\cos q_{x'}-2t_{0}\cos q_{y'})={\rm sgn}(\langle \sigma_z \rangle_g|_{q_z=\frac{\pi}{2}})$
. Second, tuning $m_z$ is equivalent to scanning $q_z$, $\langle \sigma_z \rangle_g|_{q_z=\frac{\pi}{2}}^{m_z-\Delta m_z}=\langle \sigma_z \rangle_g|_{q_z}^{m_z}$, where $\Delta m_z = 2t_z \cos q_z$. This can also be understood as scanning a virtual momentum axis $q_m=\pm\frac{k_0}{\pi}\arccos{\left(\frac{\Delta m_z}{2t_z}\right)}$. Finally we arrive at
\begin{eqnarray}
{\rm sgn}\left(\langle \sigma_z \rangle_g|_{\delta q_z}^{m_z}\right)={\rm sgn}\left(\widetilde{\langle \sigma_z \rangle}_g|_{\delta q_z=0}^{m_z-\Delta m_z}\right).
\end{eqnarray}

Therefore, the virtual slices of the BIS can be obtained by simply varying $m_z$.
Compared to the actual atom slices, the BIS on these virtual slices are thicker because the cross-over from positive polarization to negative polarization is smoothed out from the $q_z$ integral.

\begin{figure*}
\centering
\includegraphics[scale=0.425]{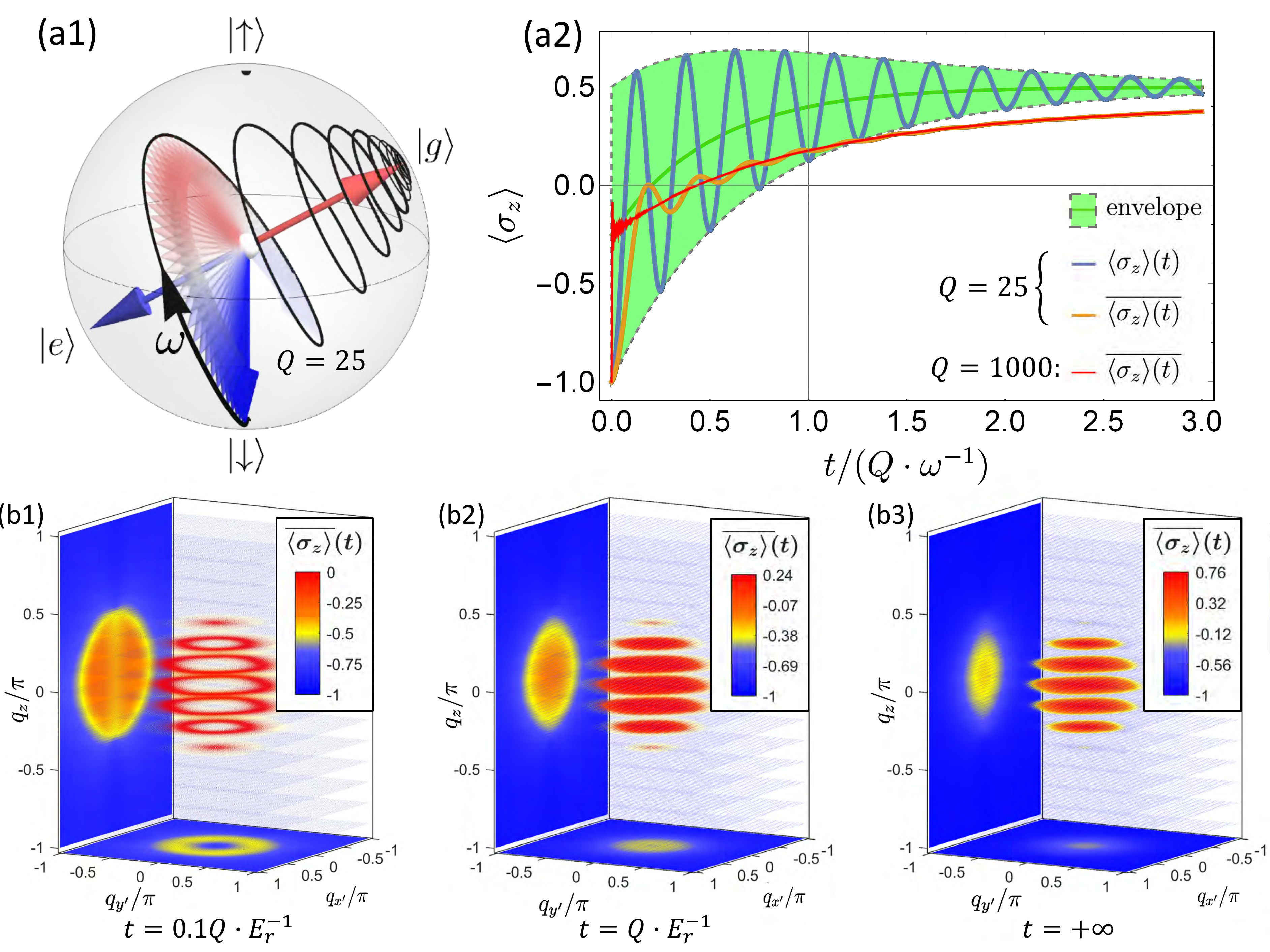}
\caption{\label{SIfig:quench_spin} {(a1-a2)} Demonstrate the evolution of the Bloch vector and spin polarization under $Q=25$. {(b1-b3)} Illustration of $\overline{\langle\sigma_z\rangle}(t)$ on the (projected) Brillouin zone versus time, with the short-time regime {(b1)} and long-time regime {(b1)} being shown.
}
\end{figure*}

\subsection{III. Measuring the 3D Weyl band: Quench Dynamics}
\label{SI:Quench}

\subsection{A. Lindblad Master Equation and Spin Oscillation}
\label{SI:Lindblad}

To study the dynamic evolution of a given initial quantum state, we use a reduced model of the Lindblad Master Equation on the tight-binding spin state of a given momentum
\begin{align}
    \Dot{\rho}(t)=-i [H,\rho(t)]+\gamma\left[L_- \rho(t) L_-^{\dagger}-\frac{1}{2}\left\{L_-^{\dagger}L_-,\rho(t)\right\}\right],
\end{align}

where we take the TB Hamiltonian in Eq.~\ref{TB_eq}: $H(\bm{q})  =\bm{h}(\bm{q})\cdot \bm{\sigma} = (m_{z}-2t_{0}\cos q_{x'}-2t_{0}\cos q_{y'}-2t_{z}\cos q_{z})\sigma_{z} +2t_{{\rm SO}}\sin q_{x'}\sigma_{x}+2t_{{\rm SO}}\sin q_{y'}\sigma_{y}$ for a given $\bm{q}$. Its corresponding eigenstates are $|1\rangle=\sin\frac{\theta}{2}|g_{\uparrow}\rangle-\cos\frac{\theta}{2}e^{i\phi}|g_{\downarrow}\rangle$ for the lower band eigenstate, and $|2\rangle=\cos\frac{\theta}{2}|g_{\uparrow}\rangle+\sin\frac{\theta}{2}e^{i\phi}|g_{\downarrow}\rangle$ for the upper band eigenstate. Accordingly, we denote the new ground-state average $\langle 1|\bm{\sigma}|1\rangle=\langle\bm{\sigma}\rangle_g$. In addition, $L_-=|1\rangle\langle2|$, and the pre-quench state $\rho(0)=|g_{\downarrow}\rangle\langle g_{\downarrow}|$.

The solution to the master equation, in the form of Bloch vector $\bm{a}(t)=\langle\bm{\sigma}\rangle(t)$ (as shown in Fig.~\ref{SIfig:quench_spin}(a1)), reads
\begin{align}
    \bm{a}(t)=R_z(\phi+\pi) R_y(-\theta)\bm{a'}(t),
\end{align}
where $R_k(\theta)$ denotes the elemental rotation matrix around $k-$axis, $\theta=\arccos(-\langle\sigma_z\rangle_g)$, $\phi=\arctan(\frac{\langle\sigma_y\rangle_g}{\langle\sigma_x\rangle_g})$, and $\bm{a'}(t)=\left(-\sin\theta e^{-\gamma t/ 2} \cos\omega t , -\sin\theta e{-\gamma t/2} \sin\omega, (1 -  \cos\theta) e^{-\gamma t} - 1\right)^{T}$ where $\omega=2|\bm{h}(\bm{q})|$. Finally we arrive at the time-evolution of the spin polarization:
\begin{equation}
    \langle\sigma_z\rangle(t) = -\left(1 - \langle\sigma_z\rangle_g^2\right)e^{-\gamma t/2}\cos{\omega t} +
   \langle\sigma_z\rangle_g\left(1-\left(\langle\sigma_z\rangle_g + 1\right)e^{-\gamma t} \right).
\end{equation}
It is in the form of a damped oscillation overlaying an exponential base function,  shown as the blue curve in Fig.~\ref{SIfig:quench_spin}(a2). The quality factor of the oscillation $Q=2\omega/\gamma$.

To average out the effect of the oscillation from the base function, we define time-averaged spin polarization as:
\begin{align}
    \overline{\langle\sigma_z\rangle(t)} = \frac{1}{t} \int_0^t \langle\sigma_z\rangle(\tau)\text{d}\tau = \langle\sigma_z\rangle_g -\frac{1}{t} \left[ \frac{2 \left(1-\langle\sigma_z\rangle_g\right) \left( \left(-\gamma \cos \omega t+2 \omega \sin \omega t\right)e^{-\frac{\gamma t}{2}}+\gamma\right)}{\gamma^2+4 \omega^2}+\frac{\langle\sigma_z\rangle_g(1+\langle\sigma_z\rangle_g) \left(1-e^{-\gamma t}\right)}{\gamma}\right] ,
\end{align}
 which damps away much more quickly (shown as the orange/red curve in Fig.~\ref{SIfig:quench_spin}(a2)), and at $Q\to \infty$, converges to \begin{equation}
     \overline{\langle\sigma_z\rangle}(t) = \langle\sigma_z\rangle_g-\frac{1}{\gamma t}\langle\sigma_z\rangle_g(1+\langle\sigma_z\rangle_g)(1-e^{-\gamma t}).
 \end{equation}

It is evident that the short-time behavior of $\overline{\langle\sigma_z\rangle}(t)$ ($\frac{4\pi}{\omega} < t < 0.1 \frac{Q}{\omega} $) differs from  long-time behavior ($t > 10 \frac{Q}{\omega} $). In the short- and long- time limit we get $\overline{\langle\sigma_z\rangle}(0) = -\langle\sigma_z\rangle_g^2$ and $\overline{\langle\sigma_z\rangle}(\infty) = \langle\sigma_z\rangle_g$, respectively. As a result, the band inversion surface (BIS) is stressed out under projection in $\overline{\langle\sigma_z\rangle}(0)$ compared to the long-time limit because in the former case spin-polarization takes negative value over the entire Brillouin zone. In Fig.~\ref{SIfig:quench_spin}(b1-b3) we illustrate how the spin-polarization and projected-spin-polarization change with respect to time evolution in the high-Q limit.  This is the advantage of the quench method: in the static spin polarization measurement, the positive and negative polarizations cancel out each other after projection, covering up the shape of the BIS; while in the the long-time behavior of $\overline{\langle\sigma_z\rangle}(t)$, the positive polarization only exist near the BIS, so the outline of the BIS is preserved after projection, as shown in Fig.~\ref{SIfig:quench_spin}(b3). From the side- and bottom- projection images, we can recreate the outline of the BIS in 3D Brillouin zone.

\begin{figure*}
\centering
\includegraphics[width=\textwidth]{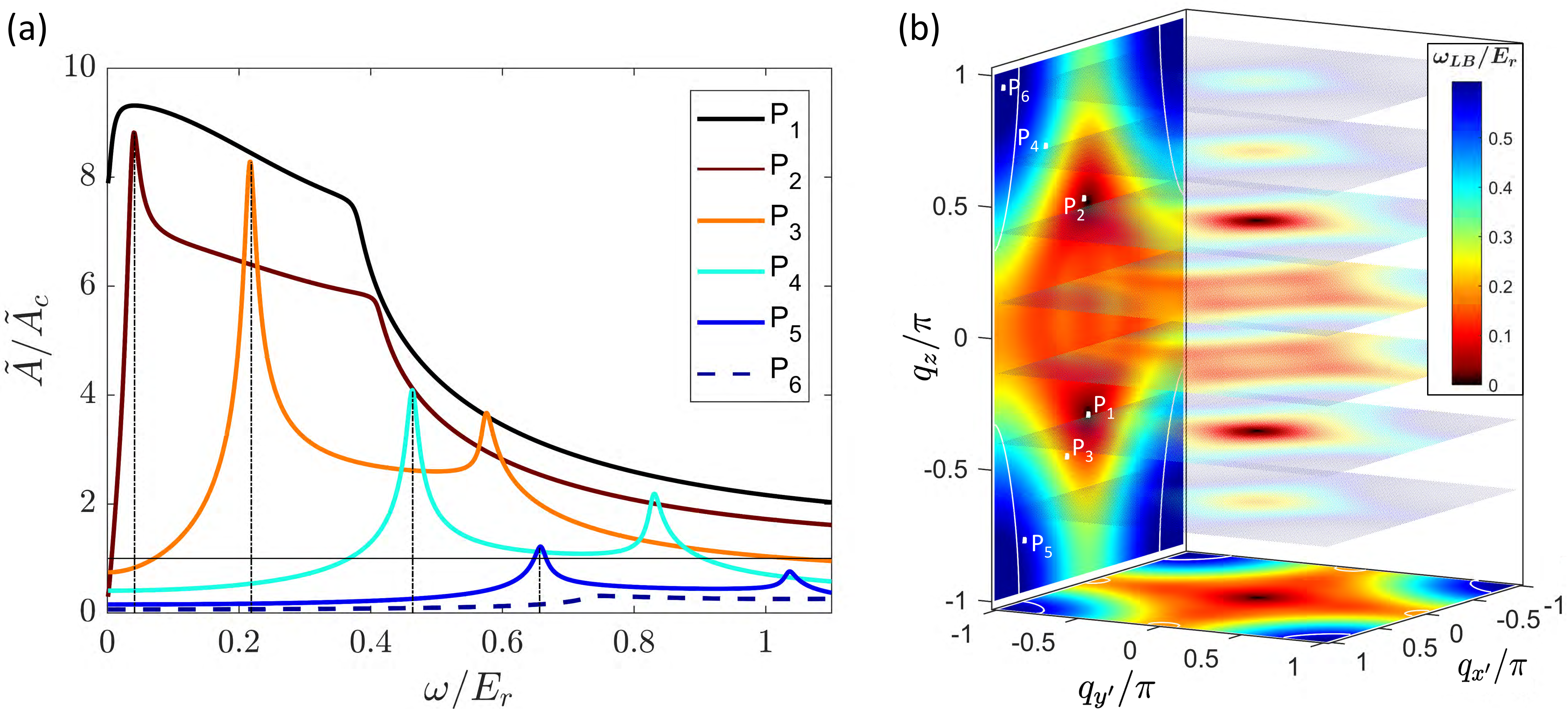}
\caption{\label{SIfig:LB}  {(a) Frequency spectrum of certain points  in the projected $\bm{k}_{x'}-\bm{k}_z$ plane at ${Q_{max}=50}$}. Each spectrum curve has two peaks, where the left(right) peak corresponds to the lower(upper) bound frequency along the projection line.  {(b) Colormap of the Larmor frequencies.} The bottom- and side- walls shows $\omega_{LB}$ in orthogonal projection.  Weyl points, being gapless, are presented as two dark spots. The white dots on the side projection corresponde to the spectra curves in {(a)}. As we move away from the projection of  Weyl points in the order of $\mathrm{P_1\rightarrow P_6}$, the signal-strengths of the peaks of the spectra curves gradually lower while their frequencies increase.
}
\end{figure*}

\subsection{B. LB Energy Spectroscopy}
\label{SI:LB}

In comparison to the static measurements, the quenching method is clearly more informative as it utilizes the time domain. Thus it offers higher resolution on the energy axis, and provides essential information on the stucture of the BIS and location of the Weyl points with the projected 2D ToF image alone. One of the ways one can make use of the real-time evolution is by marking out the lower-bound frequency of Larmor frequencies on all the points along a projeciton line - we call this method the \emph{lower-bound(LB) energy projection spectroscopy} - which roots in the linearity of Fourier transformation.

In Fig.~\ref{SIfig:LB}(b), the frequency projection images of the 3D Larmor frequency plot are shown as orthographic projections on the side- and bottom- walls, or the $q_{x'}-q_{z}$ and $q_{x'}-q_{y'}$ plane, respectively. Fig.~\ref{SIfig:LB}(b) demonstrates how to obtain $\omega_{LB}$ using the $q_{y'}$-projected imaging in the $q_{x'}-q_{z}$ plane as example. First, one can obtain a amplitude-frequency ($\tilde{A}-\omega$) curve for every point on the $q_{x'}-q_z$ plane by performing a Fourier transformation on the (projected) spin polarization oscillation data (here we introduce damping in the system so that the highest Q is $\sim$ 50).
\begin{align}
    \tilde{\mathcal{A}}(\omega)=\mathcal{F}\left[\frac{1}{2\pi}\int_0^{2\pi}\langle\sigma_z\rangle(t)] \text{d}q_{y'}\right]=\frac{1}{2\pi}\int_0^{2\pi}\mathcal{F}\left[\langle\sigma_z\rangle(t)\right] \text{d}q_{y'}.
\end{align}
What we get essentially is a summation of all the frequency-response curves along the $q_{y'}$ dimension, which results in peaks at the lowest and the highest frequencies along the integrated line. This is due to the fact that projection doesn't change frequency, but instead highlights the lowest and the highest frequency responses because the density of states at these frequencies are infinity (in theory, the two peaks of each frequency spectra curve should be infinitely tall and narrow, but the damping in the oscillation widens the peak-width). From the LB projection plots, we can straightforwardly mark out the position of each individual Weyl points, as the energy difference vanishes. Because the bands are essentially smooth, we can also infer the linear dispersion around the Weyl point from the linearity of the projected LB energy difference.

Points in some areas (separated by white lines in Fig.~\ref{SIfig:LB}(b)) on the corners of LB frequency projection plots may have uncertain $\omega_{LB}$ at the presence of noise because their LB peak signal is not obvious - such as the $\mathrm{P_6}$ curve in Fig.~\ref{SIfig:LB}(a) -  rendering these areas unresolvable. However, by cross-examining with the shape of the BIS in Fig.~\ref{SIfig:quench_spin}(b), we know that these low signal-to-noise-ratio areas are far from the BIS, and thus do not contain any Weyl points.

Note that compared to the LB frequency projection, the maximal spin polarization projection images are deduced more straightforwardly. Unlike frequency which is remained after projection, projection averages out the spin polarization.

\section{IV. Chiral Magnetic Effect}

\begin{figure*}
\centering
\includegraphics[width=\textwidth]{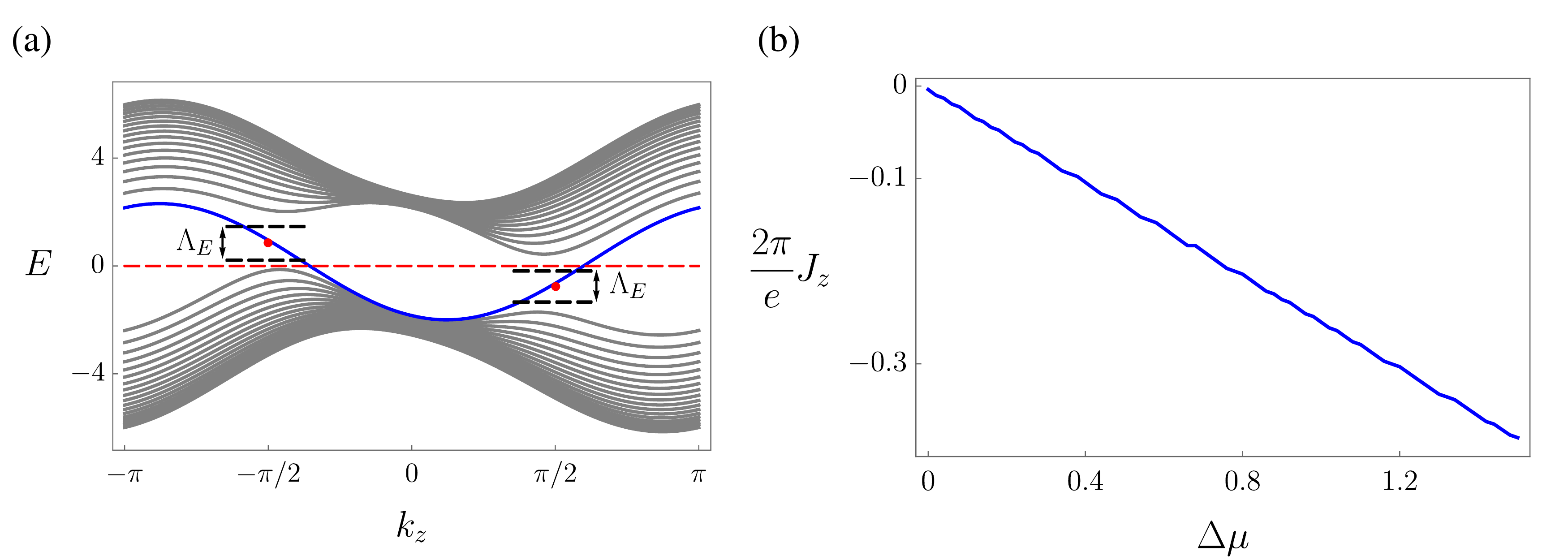}
\caption{\label{SIfig:LL}  {(a) Landau level with $t_0=t_z=t_{\rm SO}=1$, $m_z=4$, $\lambda =0.4$ and $\Phi=0.05h/e$.} The blue line labels $n=0$ Landau level, and the other gray lines are bands for $n \ne 0 $. Two red points represent Weyl points locate at $k_z=\pm \pi/2$. The red dashed line is the chemical potential, which is set zero $\mu=0$ in CME. We take the energy cutoff $\Lambda_E$ illustrated by the black dashed lines. The states within the energy domain defined by $\Lambda_E$ and below chemical potential are considered for calculating chiral current. {(b) Current of chiral anomaly.} The chemical potentials at two Weyl points are different when electric field is applied. Here we show the current changes linearly with the chemical potential imbalance.
}
\end{figure*}

The Chiral Magnetic Effect (CME),
which states that with an energy difference $2b_0$ between the Weyl points induced by breaking reflection symmetry, a topologically protected dissipationless chiral current can be generated by magnetic field even without applying electric field~\cite{Franz2013}. This effect was highly debating~\cite{CME1,CME2,CME3,CME4,CME5,CME6}, but a definite answer can be given based on the present IWSM.
In our realization of idea Weyl semimetal, the Hamiltonian used to describe CME can be written as
\begin{align}
    H(q_{y'},q_z)&= \sum_{x'} H_t(q_{y'},q_z) + H_{\rm SO}(q_{y'},q_z) + H_{\lambda}(q_{y'},q_z),
\end{align}
where $H_t$ represents the hopping term, $H_{\rm SO}$ is the SO coupling term and $H_{\lambda}$ labels a hopping term that breaks reflection symmetry along $z$ direction. The system has $(L\times L)$ sites in the $x$-$y$ plane and periodic boundary conditions, infinite sites along $z$ direction. For simplicity of numerical calculation, the Fourier transformations are performed in $y$ and $z$ directions
\begin{align}
    H_t &=  \left[ m_{z} -2t_{z}\cos q_{z} - 2t_{0}\cos ( q_{y'} - 2\pi x' \frac{\Phi}{\Phi_0} ) \right] \left(\hat{c}^\dagger_{\bold{q},\uparrow} \hat{c}_{\bold{q},\uparrow} - \hat{c}^\dagger_{\bold{q},\downarrow} \hat{c}_{\bold{q},\downarrow} \right) \notag \\
    &{} -t_0 \left( \hat{c}^\dagger_{x'+1,\uparrow} \hat{c}_{x',\uparrow} - \hat{c}^\dagger_{x'+1,\downarrow} \hat{c}_{x',\downarrow} + h.c. \right), \notag \\
   H_{\rm SO} &= 2t_{\rm SO}\sin (q_{y'}-2\pi x'\frac{\Phi}{\Phi_0}) (\hat{c}^\dagger_{q_{y'},\uparrow} \hat{c}_{q_{y'},\downarrow} + h.c.) + t_{{\rm SO}} \left( \hat{c}^\dagger_{x'+1,\uparrow} \hat{c}_{x',\downarrow} - \hat{c}^\dagger_{x'-1,\uparrow} \hat{c}_{x',\downarrow} + h.c. \right), \\
   H_{\lambda} &= -2\lambda \sin q_z \left(  \hat{c}^\dagger_{\bold{q},\uparrow} \hat{c}_{\bold{q},\uparrow} + \hat{c}^\dagger_{\bold{q},\downarrow} \hat{c}_{\bold{q},\downarrow}  \right). \notag
\end{align}
The effect of external magnetic field is included by standard Peierls substitution in the lattice model, $t \to t \exp \left[2\pi i\Phi/\Phi_0 \right]$. Here $\Phi$ is magnetic flux in each unit cell and $\Phi_0 = hc/e$ is the flux quantum. The external magnetic field is applied along $z$ direction, and the magnetic flux $\Phi$ is the same for the hopping of spin-$\uparrow$ (spin-$\downarrow$), and the flip between spin-$\uparrow$ and spin-$\downarrow$.
The reflection-symmetry-breaking $\lambda$ term in $H_\lambda$ leads to an energy shift $2b_0=2\lambda \sin q_w$ between two Weyl points.

\begin{figure}[h]
\centering
\includegraphics[width=0.62\textwidth]{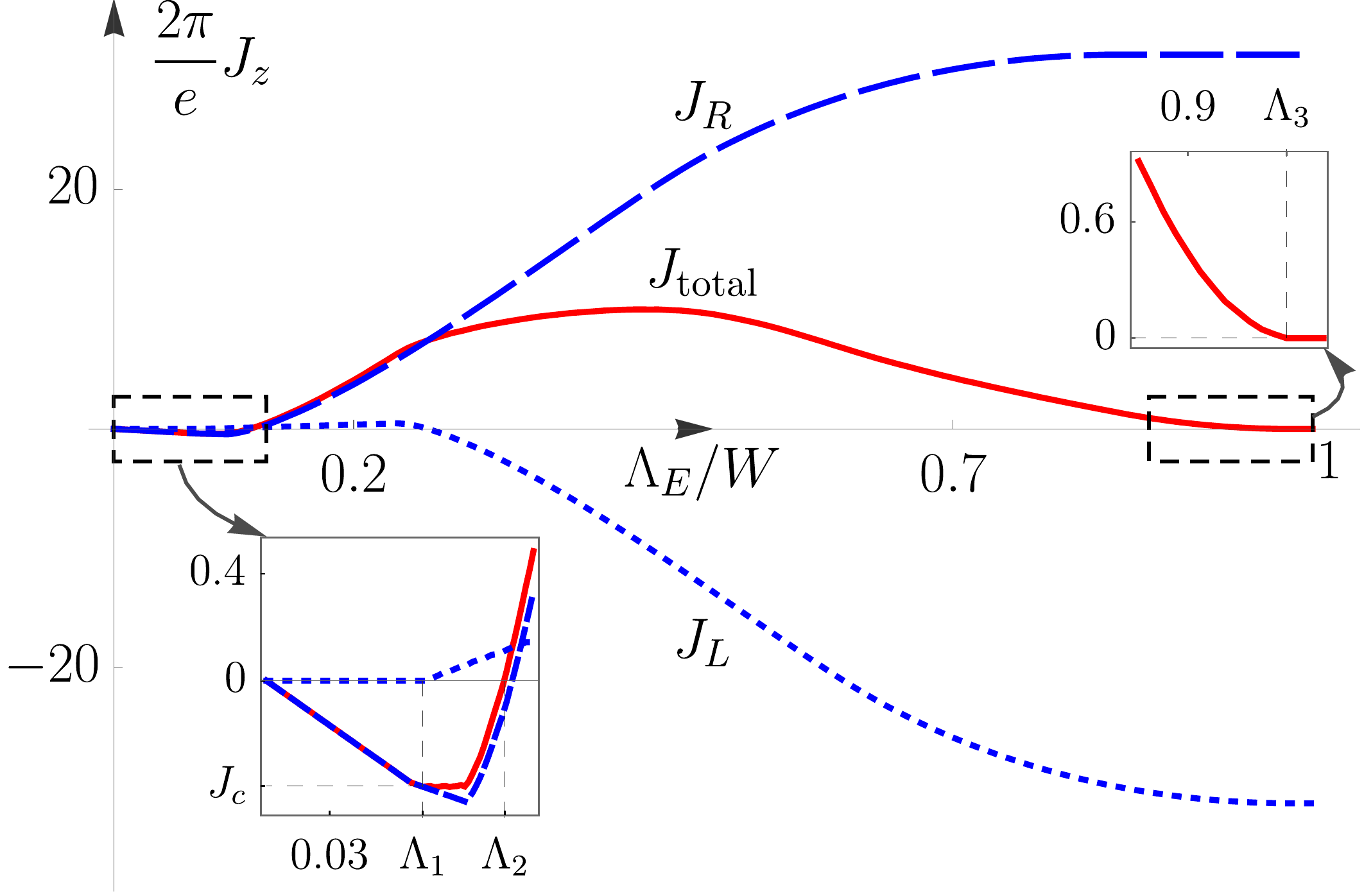}
\caption{\label{fig:CME} Examining the chiral magnetic effect. Breaking reflection symmetry leads to an energy difference $2b_0=1.6t_0$ between two Weyl points, together with a magnetic field applied along $z$ direction. The currents are plotted versus an energy cutoff $\Lambda_E$. The left (right) insert shows details for low (high) energy regime with $\Lambda_E \ll W$ ($\Lambda_E \sim W$).
The parameters taken for the IWSM are $m_z=4t_{0,z,\rm so}=4$, chemical potential $\mu=0$, and a small flux per plaquette $\Phi=0.05h/e$.}
\end{figure}

This system with magnetic field can be solved numerically, we set the parameters  as follows $m_z=4$, $t_0=t_z=t_{\rm SO}=1$, $\lambda =0.4$, $\Phi_0=h/e=1$, $\Phi =0.05\Phi_0$, $L=800$ and $\mu=0$. A pair of Weyl points are yielded at $k_z=\pm \pi/2$, and the energy shift is given via $b_0 = 0.8$. The considered magnetic field is weak so the band is not distorted much, while it is enough to generate the Landau levels. Fig.~\ref{SIfig:LL}(a) shows the Landau levels obtained in our model  with above parameters. The chiral current is given by
\begin{equation}
    J_z = -e\sum_m \int_{\rm BZ} \frac{dk_z}{2\pi} \frac{\partial \epsilon_m(k_z)}{\partial k_z} n_F[\epsilon_m(k_z)],
\end{equation}
where $\epsilon_m$ represents the energy spectrum of $m$-th band.
We show the numerical results in Fig.~\ref{fig:CME}, the chiral current $J_{\rm total}=J_L+J_R$ (red curve) versus an energy cut-off $\Lambda_E$ (in units of the half Weyl bandwidth $W$ plus $b_0$), which constrains the number of states in calculation, with $J_L$ (blue dashed) and $J_R$ (blue dotted) denoting the contributions solely from the left and right Weyl cones, respectively. In the low-energy regime with $\Lambda_E \ll W$, one has $J_{\rm total}\neq0$ and at $\Lambda_E=\Lambda_1$ (left insert) it precisely captures the CME due to the population difference in the left ($L$) and right ($R$) linear Weyl cones with ${\cal H}_w^{L,R}=2t_{\rm so}[q_{x'}\sigma_x+(q_{y'}-e/(\hbar c)x\Phi)\sigma_y]\mp2t_zq_z\sigma_z\pm b_0$. The current reverses sign for $\Lambda_E>\Lambda_2$, in which case the higher Landau levels of the bulk dominates the contribution for the intermediate-energy regime. Finally, in high-energy limit $\Lambda_E>\Lambda_3\approx W$, one always has $J_{\rm total}=0$ (right insert), showing that the dissipationless CME is absent for lattice system. This result has a simple interpretation. With magnetic field the original Weyl band splits into multiple subbands corresponding to magnetic Brillouin zones. For any equilibrially populated subband $J_L$ and $J_R$ must cancel out, independent of the subband details like reflection symmetry-breaking. The nonzero $J_{\rm total}$ in the low-energy limit is an artificial result in lattice after removing high energy contribution. Thus, the dissipationless CME cannot appear in a real crystalline solid, unless one applies an external electric field to generate a nonequilibrium distribution of fermions, which is well-known as the chiral anomaly. The current generated in chiral anomaly changes linearly with the chemical potential imbalance, as shown in Fig.~\ref{SIfig:LL}(b). On the other hand, the CME contributed solely from low-energy states can be tested with the present realization by properly loading ultracold atoms only to the states around Weyl points, which is generically not realistic for solid systems.

In the experimental, the $U(1)$ magnetic field and reflection-symmetry-breaking $\lambda$ term can be realized in experiment by modifying the optical Raman lattice configuration. First, we set the $z$-directional 1D lattice to be generated by standing-wave of wave-number $k_{V}$ which does not take part in realizing the Raman couplings and can be far detuned. Further, we apply a different running beam with wave vector $\bold{k}_0=k_1\hat e_z+k_2\hat e_{x'}$, which together with the standing waves in the $x$-$y$ plane induces the Raman couplings. The key features are two-fold. One is that the phase $\phi_z=\pi k_1/k_{V}$ is responsible for the reflection symmetry breaking in $z$ direction and the energy imbalance between the two Weyl [see Section~I, part E]. The other is that the in-plane component ($k_2$) of the wave-vector $\bold{k}$ contributes an additional phase $e^{ik_2x'}$ to the Raman potentials, which gives the $U(1)$ gauge $\bold{A}=(0,Bx',0)$ for the SO terms. Finally, the $U(1)$ gauge for the spin-conserved hopping can be created by standard laser-assisted hopping technique, e.g. through applying two additional laser beams, which could be far detuned, do not contribute to spin-flip Raman coupling but modulate spin-conserved hopping along $y'$ direction with wave-vector difference $k_2$ along $x'$ direction. 
In this way, the same $U(1)$ gauge $\bold{A}=(0,Bx',0)$ is also generated in the spin-conserved hopping. The reflection symmetry breaking and the additional $U(1)$ magnetic field along $z$ direction are then realized.

%
%

\end{document}